\documentclass[11pt]{article}
\usepackage{graphicx}
\usepackage{amsmath}
\usepackage{amssymb}
\usepackage{amsbsy}
\usepackage{amsfonts}
\usepackage{subfig}
\usepackage{amsthm}
\usepackage{textcomp}
\usepackage{mathtools}

\usepackage{etoolbox}
\usepackage[english]{babel}
\usepackage{titlesec}
\usepackage{diagbox}
\usepackage[capposition=top]{floatrow}
\usepackage{stackengine}
\usepackage{lscape}
\usepackage{array}
\usepackage{url}
\usepackage[letterpaper, margin=1.0in]{geometry}
\usepackage{fancyhdr}
\usepackage{verbatim}
\newcolumntype{M}[1]{>{\centering\arraybackslash}m{#1}}
\usepackage[flushleft]{threeparttable}
\usepackage{booktabs,caption}
\usepackage{multirow}
\usepackage{mathrsfs}
\usepackage{xcolor}
\usepackage[hidelinks]{hyperref}
\usepackage{color}
\usepackage{colortbl}
\usepackage{hhline}
\usepackage{longtable}
\usepackage{anyfontsize}
\usepackage{csquotes}
\usepackage[backend=biber, sorting=none]{biblatex}
\bibliography{Refs}
\usepackage{soul}
        \setul{0.15ex}{}
\usepackage{setspace}
\usepackage{caption,booktabs,array}

\usepackage{arydshln}

\title{A Quantitative Simulation-based Modeling Approach for College Counseling Centers}
\author{ Sohom Chaterjee$^a$, Youssef Hebaish$^a$, Lewis Ntaimo$^a$, James Deegear$^b$\\Miles Rucker$^b$, Hrayer Aprahamian$^a$\\[5pt] $^a$Department of Industrial and Systems Engineering\\$^b$ Counseling \& Psychological Services\\
Texas A\&M University,  College Station, TX 77843, United States\\}

\date{\today}
\titlespacing*{\section}
{0pt}{1.5ex plus 0.1ex minus .1ex}{0.3ex plus .1ex}
\titlespacing*{\subsection}
{0pt}{1.5ex plus 0.1ex minus .1ex}{0.3ex plus .1ex}
\titlespacing*{\subsubsection}
{0pt}{1.5ex plus 0.1ex minus .1ex}{0.3ex plus .1ex}

\begin{document}
\maketitle

\vspace{-1.1cm}
\begin{abstract}

\vspace{-0.1cm}
College counseling centers in various universities have been tasked with the important responsibility of attending to the mental health needs of their students. Owing to the unprecedented recent surge of demand for such services, college counseling centers are facing several crippling resource-level challenges. This is leading to longer wait times which limits access to critical mental health services. To address these challenges, we construct a discrete-event simulation model that captures several intricate details of their operations and provides a data-driven framework to quantify the effect of different policy changes. In contrast to existing work on this matter, which are primarily based on qualitative assessments, the considered quantitative approach has the potential to lead to key observations that can assist counseling directors in constructing a system with desirable performance. To demonstrate the benefit of the considered simulation model, we use data specific to Texas A\&M's Counseling \& Psychological Services to run a series of numerical experiments. Our results demonstrate the predictive power of the simulation model, highlight a number of key observations, and identify policy changes that result in desirable system performance.

\medskip
% \keywords{College counseling; mental health; modeling; discrete-event simulation; system performance; access time}

\end{abstract}

\renewcommand{\labelenumii}{\arabic{enumi}.\arabic{enumii}}
\renewcommand{\labelenumiii}{\arabic{enumi}.\arabic{enumii}.\arabic{enumiii}}
\renewcommand{\labelenumiv}{\arabic{enumi}.\arabic{enumii}.\arabic{enumiii}.\arabic{enumiv}}

\section{Introduction and Motivation} \label{sec:introduction}

Since the advent of the COVID-19 pandemic and its impact on our everyday lives, mental health related problems have seen a sharp increase across the population \cite{cullen2020mental,usher2020covid}, with a notable negative impact on students \cite{grubic2020student,zhai2020addressing,son2020effects}. Today, student mental health in higher education is considered one of the primary hurdles in the path to academic success. Studies have shown that students experience their first onset of mental health problems, as well as an increase in pre-existing symptoms, during their college years \cite{pedrelli2015college}. Unfortunately, over the past few decades, there has been an alarming increase in psychological issues that college students exhibit. For example, according to the 2006 National Survey of Counseling Center Directors, almost half of college-aged individuals were reported to have some kind of psychiatric disorder \cite{Blanco2008}. More recent reports from the 2019 annual survey by the Association for University and College Counseling Center Directors (AUCCCD) demonstrate that college students suffer from a multitude of mental disorders, with anxiety, depression, and stress being the most prevalent psychological issues reported by counseling centers \cite{LeViness2020}. 

To combat this worrying trend, higher education institutions have set up university Counseling and Psychological Service (CAPS) centers to provide preventive and remedial counseling to help students identify and attain personal, academic, and career goals \cite{Kitzrow2009}. However, the increasing ethnic, racial, and social diversity within the student population, as well as the changing trends in students' needs, has altered the traditional mission of these centers, with student mental health support being one of the primary services \cite{Archer}. These challenges have been exacerbated by the pandemic \cite{Lederer2021}. Campus closures and the suspension of in-person classes result in adverse psychological effects on students - such as loneliness, isolation, stress, anxiety, and depression - leading to a drastic increase in psychological needs \cite{Zhai2020}. This surge in demand has imposed additional strains on counseling centers, and the current model is unable to meet the growing mental health needs of students and is resulting in staff burnout \cite{fu2017college,ubtcu,Litam2021}, thus calling for a major transformation in the delivery of CAPS at university campuses. In this paper, we propose a quantitative framework to provide CAPS with data-driven insights and recommendations to better manage their resources and to streamline and optimize their operations. 

Counseling centers in universities are different from counseling centers that cater to the general population in two ways: First, the demand for CAPS services on college campuses follows a cyclical nature related to the student academic calendar. For instance, at the beginning of the semester, or right before major holidays (e.g., Thanksgiving, Spring Break), demand is usually sparse due to low academic pressure, but the demand starts increasing as the semester progresses. This trend is mainly due to the stress that students develop as they approach their exams and the accumulation of academic workload. This semester-based cyclic nature of demand often imposes strains on counseling centers’ resources. Second, the set of mental health disorders, and their prevalence, experienced by students is different from that of the general population. As such, CAPS facilities need to tailor their services and resources to cater to this need. Given these unique characteristics of CAPS, we focus the analysis on college counseling centers which narrows down the scope to a more manageable target population with a clear set of psychological disorders. Moreover, one of the objectives of this work is to take advantage of the recurring semester-based demand trends to help CAPS better understand their system and its performance.

College counseling centers face a myriad of challenges, which can be broadly categorized into two classes: Resource-level and patient-level challenges. Patient-level challenges pertain to the relationship between patient improvement and the treatment plan they undertake, while resource-level challenges involve resource planning and allocation. Patients have unique needs and hence are assigned tailored treatment plans which are often a combination of various treatment options (e.g., one-on-one counseling sessions, psychiatric treatment, etc.). The main challenge lies in identifying a treatment plan that yields the best possible improvement. Studies have shown that a patient's improvement tends to increase with the number of attended sessions \cite{Draper2002}. However, the study also reveals that there are no clear pathways to construct treatment plans for patients based on their specific case. Hence, there is no guarantee that the treatment plan they receive maximizes improvement. In addition, although there is a direct relation between the number of sessions and patient improvement, it may be infeasible—on a resource level—to provide every patient with as many sessions as they might need. Therefore, counseling centers also face challenges with regard to resource planning. As previously pointed out, there is a natural growth in demand due to the increasing college student population as well as their psychological issues \cite{Brunner2014, Xiao2017}. This increase in demand has negatively affected the waiting time for patients to receive care. According to the 2019 survey by AUCCCD that included 562 counseling centers, the average wait time for a first triage appointment (which we refer to as \textit{access time}) was $6.1$ days, while the average wait time for the following clinical appointment was $8.7$ days. These numbers are expected to be much higher during the pandemic. The most straightforward solution to this issue is to hire more counselors as the most frequently reported barrier to meeting demand, as reported by counseling centers directors, is understaffing issues \cite{Hu2021}. However, limited funding availability, which is the root cause of understaffing \cite{Chugani2015}, prevents the adoption of such solutions. The study in this paper focuses on addressing the resource-level challenges facing CAPS with the aim of providing recommendations that do not impact the current staffing structure and strategy for deciding on treatment plans.

To combat the aforementioned challenges, a number of attempts have been made by counseling centers to meet the surge in demand. One example is external referral of patients where students are referred to off-campus mental health providers whenever they require a higher level of specialization that cannot be provided on-campus \cite{Sharkin2012}. This is done so that counseling centers can meet their students’ needs \cite{Gallagher2012}. For example, Iarussi and Shaw \cite{Iarussi2016} propose a “Collaborative Process Model” for referral, consisting of four phases: (i) informed consent, (ii) assessment of patient’s case, (iii) collaborative decision making, and (iv) execution. The purpose of the first phase is to provide patients with all the necessary information about the scope of services provided by the counseling center and the possibility of a referral to an off-campus provider \cite{Iarussi2016}. The second phase is assessing the patient’s case during the first appointment to gather information about the patient and their needs. The third phase personalizes referral options according to the patient’s psychological needs and financial circumstances. The fourth and final phase is to initiate the referral process by connecting the patient to the new provider and following up with them \cite{Iarussi2016}. 

Owen et al. \cite{Owen2007} suggest a similar referral process that entails meeting the patient and then following up with them. Their study showed that $58\%$ of patients at a university counseling center were successful in connecting with an off-campus provider. Although such models might result in more successful referral cases due to the personalization nature of the process, they have some issues concerning resource allocation and functionality. First, the patient’s involvement in the process increases the likelihood of successful referral (even if it is not optimal to do so); however, this claim is not substantiated by empirical evidence that supports its validity. Second, such models do not tackle one of the significant challenges that counseling centers face, access time \cite{LeViness2020}. Because the initiation of the process begins with the first phase during the first appointment with the patient, such models do not affect the expected access time. In addition, such studies do not provide any information regarding the impact external referrals have on the overall system performance. Another example of an attempt made by counseling centers is to set an upper bound on the number of sessions that a student can utilize. This decision was driven by the fact that a considerable number of students end up utilizing a large number of treatment sessions which prevents other students from accessing mental health services. While such a policy has the potential to reduce the load on CAPS and increase access to care, it is still not clear how such a change will impact the overall performance. Moreover, no framework currently exists to help CAPS facilities determine an appropriate maximum number of sessions.

Due to the ever-growing complexities of healthcare systems, quantitative analytical methods that attempt to enhance services and reduce costs have been receiving significant attention \cite{Rais2011}. An example is optimization which is one of the most commonly used operations research tools in healthcare settings \cite{Batun2013}. While the field is rich with a vast amount of literature, studies that focus on counseling centers in universities are scarce. For instance, Began and Queyranne \cite{Begen2011} discuss an efficient method to solve a single-server appointment scheduling problem. Their objective is to design an optimal schedule for a given sequence of jobs with the assumption that job durations are integer random values. Denton and Gupta \cite{Denton2010} also consider a single-server scheduling problem using a two-stage stochastic linear program, solving it using a modified L-shaped algorithm, with the objective of minimizing waiting time and service overtime. Although these optimization models are robust and provide a mathematical basis for scheduling policies, they exhibit difficulty capturing the complexities of most healthcare settings. For instance, analytical models are often limited to Exponential or Erlang service times \cite{Cayirli2003}. In addition, as seen in \cite{Begen2011,Denton2010}, such optimization models are often formulated for a single server for simplicity. While multi-server approaches have been considered (e.g., \cite{Alaeddini2011,Granja2014,Liu2010,Sickinger2009}), existing studies often impose the unrealistic assumption of identical servers since the incorporation of nonidentical servers leads to analytical intractability. 

Another commonly utilized analytical approach in healthcare settings is discrete-event simulation (DES). DES is a technique used to study discrete-event dynamic systems by generating ``sample paths" that mimic a system's behavior \cite{Fishman2001}. A major advantage that DES models offer is the ability to capture real-life complexities that healthcare systems exhibit (e.g., patient cancellations and no-shows, dynamic schedule adjustments, etc.). Such complicating factors, which are often neglected in optimization models for tractability reasons, can play a crucial role in governing the performance of the system and hence should not be ignored. In addition, a DES model is a valuable tool that allows conducting various what-if experiments, especially when real-life implementation are expensive or not possible to conduct. Thus, DES models are a powerful tool to study scheduling problems in complex, highly stochastic, systems such as our setting. Although numerous DES models have been developed for various healthcare settings \cite{Fishman2001,Ala2020,Klassen2009,Zhang2018},  simulations of counseling centers are not abundantly found in the literature, leaving a critical gap.

In this paper, given the aforementioned distinct advantages of simulation models, we build a DES model to study and analyze counseling centers' performance. The main objective is to provide a quantitative framework to help answer fundamental questions about CAPS's operations so as to propose data-driven solutions and recommendations that improve system performance. By utilizing this DES model, this paper aims to address three key research questions: (i) what is the effect of external referrals on the performance of the system and what proportion of external referrals is needed to attain desirable performance? (ii) what is the impact of a maximum session policy on the system performance and what upper bound provides good performance? (iii) how does the schedule topology impact system performance and what topologies lead to favorable performance? Addressing these research questions is of importance as doing so can result in significant performance improvement. This, in turn, will positively impact students by increasing access to critical mental health care. To measure system performance, the existing literature on simulation models that tackle healthcare related scheduling operations often use waiting time as the Key Performance Indicator (KPI), e.g., \cite{Burns2022, Valente2009, Aboueljinane2014, Santi2009}. In this study, we follow suit but use two KPIs which are the average wait time for first appointment (access time) and the average wait time of crisis patients (referred to as \textit{crisis time}). These KPIs were chosen based on lengthy discussions with our collaborator Texas A\&M (TAMU) Counseling \& Psychological Services. In fact, our close collaboration with CAPS provides domain expertise and real-life data of patients’ arrival and service times, attended sessions, and information on cancellation and no-show. This is of great value as it enabled us to construct and validate a simulation model that is realistic \cite{Altiok2007}. In summary, the contributions of this paper are two-fold: First, to the best of our knowledge, this paper is the first to present a DES model that is specifically tailored to CAPS. The model serves as a quantitative platform to help CAPS facilities identify ways to improve system performance. The simulation model will be publicly available for other counseling centers to benefit from it. Second, we conduct a case study using data specific to TAMU CAPS. Our simulation results provide data-driven insights on the impact of external referrals, maximum session policy, and scheduling topology on system performance.

The remainder of this paper is organized as follows: Section \ref{sec:scheduling} provides an overview of the scheduling system implemented at CAPS centers. Then, Section \ref{sec:flow} presents a high-level description of the operational flow at CAPS centers. Section \ref{sec: Sim} details the different elements of the discrete-event simulation model and their implementation. Section \ref{sec: casestudy} discusses findings and results from a series of numerical experiments conducted on TAMU CAPS. Lastly, Section \ref{sec: conclusion} concludes the paper and provides future research directions.

\section{Overview of CAPS Scheduling System}
\label{sec:scheduling}

College counseling centers are equipped with a wide range of employees, including clinical staff, trainees, and supporting staff. The clinical staff (hereafter collectively referred to as counselors) includes licensed psychologists and psychiatrists who provide counseling, assessment, and treatment for behavioral, emotional, and mental disorders. While all counselors are trained as generalists, some have specialty in specific domains (e.g., psychiatric treatment), which makes them particularly suited to handle certain patients. Collectively, counselors represent the majority of the employee-base at CAPS and are the key operators of the facility. Consequently, appropriately managing their time and effort (which is the responsibility of the director) is key to achieve a good performing system that effectively meets student demand. The duties and time commitment of counselors can be broadly categorized into four main components (referred to as service types). The first three service types correspond to serving three different types of patients: (i) first-time, (ii) ongoing, and  (iii) crisis. As the name implies, first-time patients are new patients that will be seen for the first time. Ongoing patients represent patients that have been previously seen by a counselor and that are now part of an ongoing treatment plan. Crisis patients, on the other hand, are patients that are going through a mental health emergency and that need immediate attention. The fourth and last service type is referred to as ``other'' which includes a host of activities (e.g., organizing and conducting workshops, service activities) that counselors are responsible for. The allocation of the counselors' time across these four service types plays a key factor in characterizing the overall performance of the system.

\begin{figure}
\caption{Example of a schedule topology for two counselors across two days.}
\centering
\vspace{-0.3cm}
\includegraphics[trim=0.5cm 6cm 0cm 6cm, clip, width=178mm]{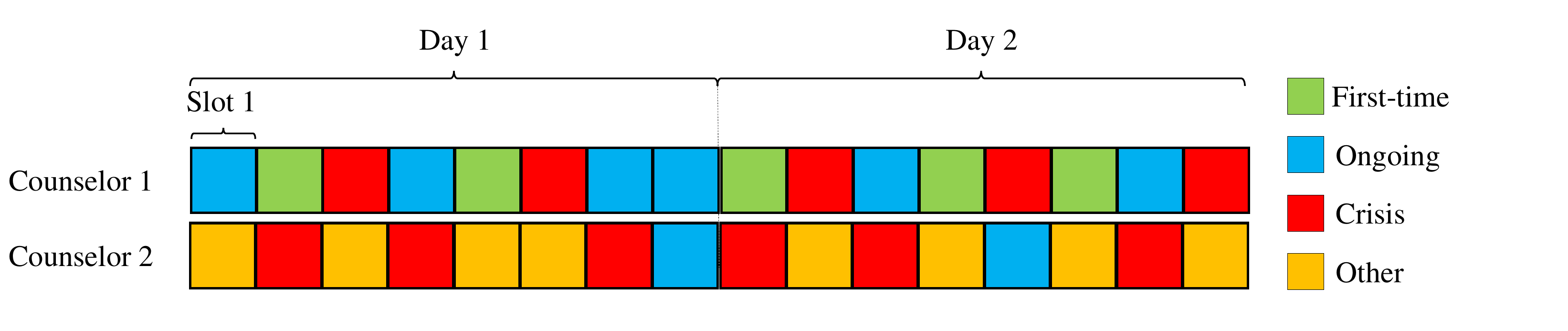}
\vspace{-1.0cm}
\label{fig:First_stage}
\end{figure}

In an effort to have a well-planned semester cycle, it is typical for CAPS directors to construct and commit to a \textit{master schedule} plan at the beginning of each semester. This plan outlines, for each counselor, a distribution of time commitment for each of the aforementioned four service types that spans the entire semester. This distribution is referred to as a \textit{schedule topology}. Figure \ref{fig:First_stage} provides an example of a schedule topology for two counselors across two days. In this example, a day is made up of eight slots each of which is allocated to one of the four service types. For example, counselor 1 has a higher emphasis for serving ongoing patients with slots 1, 4, 7, and 8 of day 1 and slots 3 and 7 of day 2 allocated to serving ongoing patients. Counselor 2, on the other hand, has a higher commitment to the ``other'' service type with slots 1, 3, 5, and 6 of day 1 and slots 2, 4, 6, and 8 of day 2 all allocated to ``other''. Establishing such a schedule topology at the beginning of each semester cycle is of great importance to CAPS directors because it provides three distinct advantages: First, doing so leads to a structured, and transparent, mechanism that greatly facilitates the effective management of a large number of counselors. Second, it allows directors to craft schedules that take advantage of specific strengths of counselors. For example, if a specific counselor has been trained to effectively handle crisis patients, then the director may lean towards a schedule topology that allocates more ``crisis'' service types to that counselor. Third, a schedule topology provides directors a high-level picture of all operations which allows them to devise schedules with sought after specifications. For example, a schedule topology can be used to ensure a fair and equitable distribution of workload among counselors. Alternatively, a schedule topology can be used to make sure certain thresholds (with respect to time committed to each service type) are met. For instance, a director may want to ensure that at least 30\% of a counselor's time is committed to serving first-time patients. 

Once a topology is committed to, arriving patients throughout the semester are scheduled in a manner that follows the selected topology. For example, if a first-time patient enters the system, then they can only be scheduled to ``first-time'' slots (i.e., time slots that have been allocated to handle first-time patients). Clearly, the schedule topology will play a major role in determining the overall performance of the system (measured through the two KPIs discussed in Section \ref{sec:introduction}). For example, a schedule topology that does not commit enough time to vulnerable crisis patients will inevitably lead to high average wait times for crisis patients. This is undesirable as it can potentially lead to detrimental consequences such as the patient being unsafe to themselves or to people around them. Similarly, if the schedule topology does not dedicate enough first-time slots, then this will increase the average access time (i.e., the wait time for a first triage appointment) of patients. Again, this is undesirable as it might deter students from accessing critical mental health services. As such, determining an appropriate schedule topology that provides favorable system performance is of utmost importance. However, identifying such a topology is extremely challenging as the decision needs to be made under high levels of uncertainty. This is the case because the schedule topology is set up at the beginning of the semester prior to realizing any of the patient arrivals. Part of the objective of this paper is to provide a simulation-based mechanism to quantify the performance of a given topology, especially ones that take advantage of certain trends that are specific to CAPS. For example, owing to the cyclical nature of demand, a good performing topology is expected to contain a high number of slots dedicated to first-time patients right before final exams. Similarly, during major holidays, a good performing topology might dedicate more slots to the service type ``other''. Such a tool is extremely valuable to CAPS directors as it can assist them in identifying good performing topologies that are specifically tailored to their needs.

\section{Overview of CAPS Operational Flow}
\label{sec:flow}

This section aims to provide a high-level overview of the operational flow at CAPS and the sequence of processes (referred to as \textit{paths}) that patients go through when entering the system. A patient's path is heavily governed by its type (i.e., first-time, ongoing, crisis). In general, there are two ways by which patients can enter the system: First, if the arrival is a first-time patient, then students must book a first-time appointment by either using an online portal or, if they need additional assistance, by visiting CAPS in-person. While there are some operational differences between these two types of booking approaches (e.g., an in-person first-time appointment booking requires students to complete a survey-based assessment report), the differences do not involve counselors' time and do not impact the two considered KPIs. Consequently, in this study we do not differentiate between these two first-time appointment booking methods. Note that the time between a patient's request for a first-time appointment and the actual appointment time is the access time (which is one of our two KPIs). Second, if the arrival is a crisis patient, then the process is simpler as no appointments are required and the patient can simply walk in during service hours and request a crisis meeting. Given the urgency of the situation, it is key for CAPS to provide the service as soon as possible. 
Of course, to be able to achieve such a service level, the schedule topology discussed in Section \ref{sec:scheduling} must allocate sufficient slots for crisis patients. This, however, comes at a cost of having fewer first-time and ongoing slots which negatively impacts access time. Such tradeoffs between the two KPIs further highlight the difficulty in identifying good performing topologies. For both system entry types, counselors are selected in a probabilistic manner based on their availability; however, preferences are given to counselors that are available earlier.

Upon completing the first session (whether as a first-time or crisis patient), counselors determine the best course of action for their patients. Consequently, the subsequent path that students end up going through heavily depends on their specific characteristics. In fact, the set of all possible patient paths is quite diverse. For example, one of the most common paths is based on a traditional one-on-one treatment cycle. In such a path, the patient is assigned to a specific counselor that they see on a periodic basis (e.g., weekly). The frequency of the meetings depends on several factors including, but not limited to, patient needs, counselor availability, and scheduling restrictions. The student continues attending one-on-one sessions until the counselor determines that the patient no longer needs treatment. The number of sessions that students end up attending can thus vary greatly; however, with the use of historical record data, it is possible to identify an appropriate distribution that replicates the current behavior. Replicating the current treatment behavior brings us back to the fact that this paper focuses on the resource-level challenges facing CAPS. That is, the aim is to analyze the system in a manner that does not impact the current strategy for deciding on treatment plans.

While the structure of treatment plans certainly plays a major role in characterizing the performance of the system, such patient-level challenges are beyond the scope of the present paper. Another example of a student path involves psychiatric treatment in which a student is assigned (in addition to a counselor) to a psychiatrist. In such a path, students end up attending two independent streams of sessions (one with a counselor and another with a psychiatrist). While such paths are much rarer than the aforementioned conventional one-on-one path, it is still important to model them as the number of available psychiatrists at CAPS is often severely limited (owing to the high costs associated with hiring licensed psychiatrists). In both of these paths, the type of the patient will transition from ``first-time'' to ``ongoing''. This transition is not always guaranteed to occur for all paths (e.g., patient paths that involve referring the student to off-campus mental health providers). While it is not possible to concisely go over all possible patient paths, numerous lengthy discussions with TAMU CAPS enabled us to comprehensively identify and factor in all possible paths that students can take.

\begin{figure}[t!]
\caption{Schematic summary of the operational flow at CAPS.} 
\centering
\includegraphics[trim=1cm 3.5cm 1cm 4.5cm, clip, width=165mm]{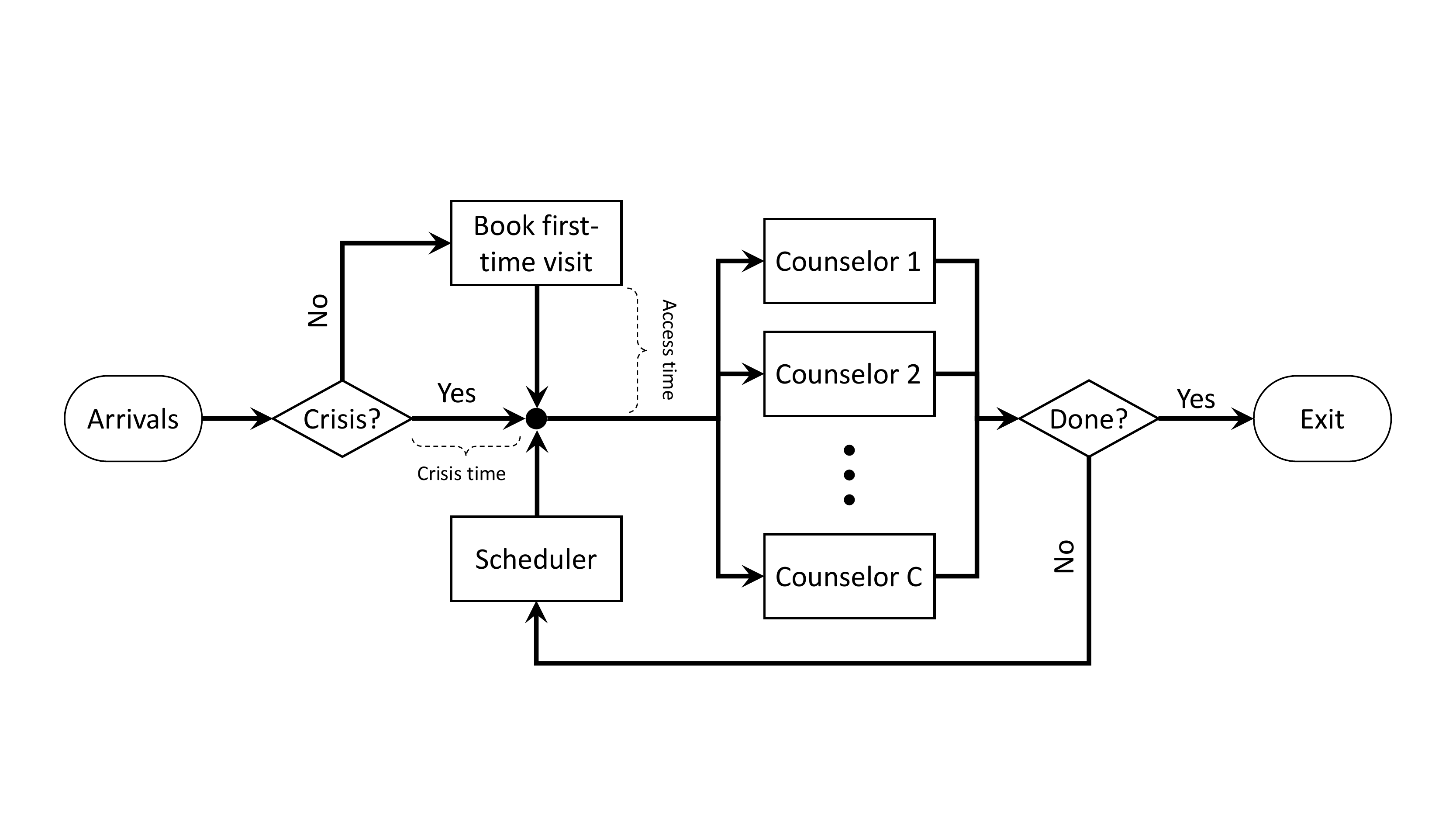}
\label{fig:CAPSflow}
\vspace{-0.6cm}
\end{figure}

Figure \ref{fig:CAPSflow} provides a high-level schematic summary of the operational flow at CAPS. Starting from the left, patients arrive and enter the system. If the arriving patient type is ``first-time'', then an appointment for the first-time visit is booked and, when the appointment time is attained, the patient proceeds to CAPS. As shown in the figure, the difference between the request and appointment times is the access time. If, on the other hand, the arriving patient type is ``crisis'', then the patient directly proceeds to CAPS and the time needed to serve the patient is the crisis time. Once the patient arrives at CAPS they will be seen by a counselor and at the end of session it will be decided whether the treatment for the patient needs to be continued or not. If yes, the patient is sent to the ``scheduler'' where new appointments are booked based on the patient type ``ongoing''. The patient will then wait until the selected appointment time is attained, after which they will head back to CAPS. Alternatively, if it is deemed by the counselor at the end of the session that no further treatment is required (or if the student is referred to off-campus mental health providers), the patient proceeds to exiting the system.

The operational flow at CAPS is subject to high levels of uncertainties and variations that arise from numerous sources. For instance, as previously discussed, the paths that students end up going through varies substantially by a number of random and uncontrollable factors. These include the patient type, the specific characteristics and needs of the patient, and the counselors' decision for determining an appropriate treatment plan. In addition to the variations in the patients' paths, several other sources of randomness exist. For example, students might cancel their appointments. In such a case, the time slot that was previously booked needs to be freed up. It is also possible for students not to show up to an appointment (referred to as \textit{no-shows}). In this case it is difficult to make the time slot available for immediate booking. Therefore, in case of no-shows, time slots are often reallocated to serve crisis patients. Other examples of uncertainty include time-varying arrival rates (recall the cyclical demand pattern), counselor downtime, and random services times. These uncertainties, when compounded, result in a highly stochastic system that is challenging to analyze and understand. However, DES models are particularly suited to handle such systems and this was one of the primary motivations for us to consider such models in the first place.

\section{The Simulation Model} \label{sec: Sim}

The discrete-event simulation model was implemented in Simio\textsuperscript{\small\textregistered} which is a popular commercial simulation software that has a set of expansive features that allow for modeling complex simulation processes \cite{simio}. Our software choice is motivated by three factors: First, Simio provides several layers of customization which facilitates the incorporation of all the necessary factors into the simulation. In particular, the ``Add-on Processes’’ feature available in Simio provides users the flexibility to add custom logic specific to the application at hand. Second, Simio has a built-in feature that allows it to read from, and dynamically modify, external files while the simulation is running. As will become clear, this feature is crucial for the successful implementation of the schedule topology discussed in Section \ref{sec:scheduling}.  Lastly, Simio provides a relatively simple drag-and-drop user-friendly interface. This ease of use not only facilitates the process of constructing the simulation, but it also allows non-domain experts (e.g., CAPS directors) to more easily visualize, interpret, and use the simulation model. The latter is of particular importance because it provides CAPS with a long-term and sustainable tool to effectively manage their operations.

\begin{table}[t!] 
  \begin{threeparttable}
  \small
    \caption{Example of the tabular format for the schedule topology presented in Figure \ref{fig:First_stage}.}
     \begin{tabular}{cccccccc}
        \toprule
        &&&\multicolumn{2}{c}{Counselor 1}&&\multicolumn{2}{c}{Counselor 2}\\
        \cline{4-5}
        \cline{7-8}
        Day&Slot&&Service type$^\dagger$&Booked?$^\ddagger$&&Service type$^\dagger$&Booked?$^\ddagger$\\
        \hline
        \multirow{8}{*}{1}&1&&1&0&&3&0\\
        &2&&0&0&&2&0\\
        &3&&2&0&&3&0\\
        &4&&1&0&&2&0\\
        &5&&0&0&&3&0\\
        &6&&2&0&&3&0\\
        &7&&1&0&&2&0\\
        &8&&1&0&&1&0\\[3pt]
        \hdashline\\[-6pt]
        \multirow{8}{*}{2}&9&&0&0&&2&0\\
        &10&&2&0&&3&0\\
        &11&&1&0&&2&0\\
        &12&&0&0&&3&0\\
        &13&&2&0&&1&0\\
        &14&&0&0&&3&0\\
        &15&&1&0&&2&0\\
        &16&&2&0&&3&0\\  
        \bottomrule
     \end{tabular}
\label{tab:topology}
\begin{tablenotes}
\item[$\dagger$] ``First-time''=0, ``Ongoing''=1, ``Crisis''=2, and ``Other''=3.
\item[$\ddagger$] ``No''=0 and ``Yes''=1.
\end{tablenotes}
  \end{threeparttable}
\end{table}

Recall that part of the objective of the simulation is to quantify the performance of the system for a given schedule topology. Given this, we first discuss how a schedule topology is defined and embedded within the Simio framework. We characterize a schedule topology by an external excel file that stores all the necessary information in an easily readable tabular format. Table \ref{tab:topology} provides an example of this tabular format for the schedule topology presented in Figure \ref{fig:First_stage}. Each row of the table represents a time slot. For example, the first 8 rows of the table represent the 8 time slots of day 1 while the next 8 rows represent the 8 time slots of day 2. The first two columns of the table provide information regarding the day number and slot number. The remaining columns are partitioned based on the counselors with each counselor having two columns: ``Service type'' and ``Booked?''. The ``Service type'' column provides information about the specific type of service allocated to that particular counselor and time slot. The values 0, 1, 2, and 3 respectively correspond to the service types ``first-time'', ``ongoing'', ``crisis'', and ``other''. For example, looking at row 1 of the table (i.e., the first time slot of day 1), counselor 1 has a value of 1 (i.e., ``ongoing'') while counselor 2 has a value of 3 (i.e., ``other''). The column ``Booked?'', on the other hand, provides information regarding the current availability of that particular counselor and time slot. A value of 0 represents a situation in which the slot is not booked (i.e., it is available for booking) while a value of 1 indicates that this slot has already been booked and hence is not available for future patients. Notice that the example provided in Table \ref{tab:topology} has all of the ``Booked?'' columns set to zero for all counselors. This particular instance, which represents a situation in which all counselors are available, is typically used as the initial table in the simulation. This is the case because at the beginning of a semester cycle (prior to realizing any of the patient arrivals) counselors have not yet been booked and hence are available. The considered approach, however, is flexible enough to consider other initial tables where certain counselor/slot pairs are already booked. Such situations may arise, for instance, when treatment cycles of patients span across multiple semesters.

The external excel file discussed in the previous paragraph plays a pivotal role in governing the overall performance of the simulation. In addition to the schedule topology, the file also embeds critical information regarding the current scheduling status of the system. As the simulation progresses, this external file needs to be dynamically updated to reflect recent bookings and modifications. For example, if a crisis patient arrives at time $0$, then (based on the schedule provided in Table \ref{tab:topology}) the nearest available crisis slot is given by slot 2 of counselor 2. In such a case, a reservation will be made for the student to attend that session and the value of the corresponding ``Booked?'' column must be updated from 0 to 1. Doing such a change on-the-fly is critical as it prevents future patients from accessing the already reserved slot. Another example is when a student cancels an appointment. Here, the value in the column ``Booked?'' must be updated from 1 to 0 to make the slot available for future patients. Since topology information is embedded within the tabular form, all resulting scheduling processes are guaranteed to adhere to the selected schedule topology. It is worth highlighting that in certain circumstances the column ``Service type'' may be dynamically updated as well. For example, if a no-show occurs, then the service type of the corresponding slot is modified to 2 to make the slot available to crisis patients. While this inherently modifies the selected topology, these modifications may be set into place by CAPS directors. Such specific complexities are extremely difficult to consider using other modeling approaches (e.g., optimization), which further motivates the use of a simulation-based model.

Having discussed how a schedule topology is defined and embedded within the Simio framework, we now turn our attention to simulating the operational flow of CAPS  (discussed in Section \ref{sec:flow}) for a given schedule topology. Primarily, the simulation model has two main elements: Counselors and patients. Counselors are modeled as servers while patients are represented as discrete entities that enter the system seeking service. The general sequence of events is as follows: Patients enter the system according to a non-stationary process that is obtained from historical record data. This arrival process will embed critical cyclical demand patterns that arise in practice. Arriving patients are then categorized to two types: Regular and crisis. Regular patients are patients that do not encounter any crisis attacks throughout their treatment cycle. In contrast, crisis patients are patients that will encounter at least one crisis attack during their course at CAPS. The crisis attacks can either occur at the beginning of their treatment cycle (i.e., they enter the system via a crisis session) or they can occur in between one-on-one sessions. Depending on availability, whenever a patient experiences a crisis attack, the counselor assigned may be different from any previous crisis counselor they have visited, or their regular counselor with which they have periodic one-on-one sessions. All of these crisis attacks will contribute to the overall average waiting time of crisis patients. The patients' type will heavily impact the path that they will end up going through (as observed from historical record data), which is why these two categories of patients were considered in the first place. The proportion of regular and crisis patients is determined from historical data.

Upon arrival, each patient entity is assigned certain key features depending on their type. For example, some features for regular patients include the total number of sessions, and no-show and cancellation probabilities. Crisis patients, on the other hand, are assigned additional features such as crisis attack probabilities. Differentiating patients allows us to customize the features based on their type. For instance, an analysis of the data reveals that crisis patients end up requiring more sessions than regular patients. Also, the no-show and cancellation probabilities between the two types of patients are different. These features, which are not necessarily deterministic, will be obtained by conducting an input analysis of the data. For instance, for the total number of sessions of regular patients, an input analysis on a subset of the data that only includes regular patients will generate a probability mass function. This distribution can then be used generate random realizations for the total number of sessions which are then assigned to newly arriving regular patients. Note that this procedure is essentially replicating the current treatment behavior at CAPS. Again, this brings us back to the fact that this paper focuses on the resource-level challenges facing CAPS with the aim of analyzing the system in a manner that does not impact the current strategy for deciding on treatment plans. Collectively, the features that are assigned to a patient will heavily govern the path that the patient ends up going through. 

Once the initial set of features has been assigned for each patient, the patient enters a scheduler station in which they will be assigned the time and counselor for their first-time session. This assignment process heavily depends on the schedule topology and the current availability of the counselors, which are recorded in the external file discussed at the beginning of this section. To achieve the assignment, we first note that the service type of a triage session can either be ``first-time'' or ``crisis'', and the scheduling process is different between the two. In particular, since a crisis is an emergency which needs urgent attention, patients who require a crisis session (which are patients that show up to CAPS without any prior reservation) are allotted the earliest available crisis time slot among all the counselors. In contrast, patients seeking non-crisis sessions (which must occur through a reservation), might be inclined to meet a specific counselor for their triage session, who might not necessarily have the earliest available slot among all counselors. To incorporate such patient-specific preferences into the simulation, we construct a custom probability distribution for the patient to choose an appropriate counselor based on the earliest availability of all counselors at the time of scheduling. The structure of the distribution is based on the fact that patients will have a higher preference for counselors with earlier availability, and thus the custom distribution assigns a higher probability to such counselors.

Once the first-time session is booked, the external excel file is updated to reflect this change and the patient waits until the assigned session time and then visits their assigned counselor. At the end of the session, a number of checks are performed which are based on the features that have been assigned to the patient. For example, if (according to the features) additional sessions are not required, then the patient will exit the system. Similarly, if it is deemed that the patient will be referred to an off-campus mental health provider, then the patient will exit the system. This referral decision is probabilistic and depends on the patient type. Alternatively, if additional sessions are required, then the patient is sent back to the scheduler to determine an appropriate time-slot for their next session. The appointment time of the next session depends on several factors such as counselor preferences, patient needs, and  schedule availability. For example, based on an analysis of the data, the vast majority of counselors prefer to see patients on a weekly or bi-weekly basis; however, the data also reveals significant deviations from this behavior.

To accurately emulate this random behavior we consider a two-step approach: First, a random number is generated that represents when (in weeks) the patient should ideally be seen. For example, a counselor might decide to want to see the patient one week from now. This probability distribution is determined from historical data (see, for example, Figure \ref{fig:timebtwsessions} in Section \ref{sec: IA}). Generally, a decreasing trend is often observed, that is, more counselors opt to see patients sooner rather than later. Owing to schedule availability, however, a counselor's ideal choice may not be feasible. For instance, it might not be possible to schedule an appointment exactly one week from now. Consequently, in the second step, we run a procedure that searches the schedule for available time slots. This search procedure starts on the counselor's ideal day (which was defined in the first step), and if an available time slot (with an appropriate service type) is found then a reservation is made. If, on the other hand, an available time slot is not found on the counselor's ideal day, the search algorithm attempts to identify slots ``around'' the ideal day. This is achieved through a sequential procedure in which the algorithm searches for available time slots within one day, two days, etc. This is repeated until a time slot is identified. Note that a consequence of this procedure is that the final reserved time slot might substantially deviate from a counselor's ideal time. This behavior is consistent with the data (e.g., in Figure \ref{fig:timebtwsessions} the bars around the weekly peaks represent such deviations). Once the next session is scheduled, the external excel file is updated to reflect this change and the process repeats (i.e., the patient waits until the assigned session time and then visits their assigned counselor).

\section{Case Study: TAMU CAPS} \label{sec: casestudy}

In this section, we use the established DES model to conduct a case study on the specific operations of Texas A\&M's counseling Center (TAMU CAPS). In Fall 2021, TAMU reported a total student enrollment of 72,982 (92\% of which enrolled in the main campus in College Station) making it one of the largest universities (by enrollment) in the entire nation \cite{TAMU}. Owing to the large student population, TAMU CAPS has been particularly affected by the unprecedented surge in demand for counseling services. This, in turn, makes TAMU CAPS an ideal candidate for such an analysis as it is in need of identifying strategies that increase access to mental health services. The main objective of this study is to utilize the DES model to quantify the impact of certain key parameters on the overall operations of CAPS. In particular, we perform a series of experiments to investigate the impact, on the KPIs defined in Section \ref{sec:introduction}, of: (i) varying the proportion of external referrals, (ii) imposing a maximum session limit, and (iii) the structure of the schedule topology. In what follows, we first provide a detailed description of the data sources used in this study (Section \ref{sec:datades}). Next, we discuss the main input parameters used in the DES model and describe the methods used to obtain these parameters from the data sources (Section \ref{sec: IA}). We then validate our DES model by running an experiment with the current operating parameters and comparing its performance to the reported performance of TAMU CAPS (Section \ref{sec:validation}). Finally, we run the aforementioned series of simulation experiments with varying operating parameters and perform sensitivity analysis for each of our KPIs (Section \ref{sec:experiments}).

\subsection{Time Horizon and Data Description} \label{sec:datades}

We run the numerical experiment for the Fall $2019$ semester which spans across 18 weeks starting from  09/2/2019 up until 12/30/2019. While the semester officially ends a couple of weeks prior to 12/30/2019, we decided to include this time period within our simulation experiment as the data revealed student arrivals during this time period. The Fall 2019 semester was chosen in order to avoid any anomalies in the data during the onset of COVID-19. However, it is possible to run the numerical experiments for later semesters to more rigorously quantify the impact of COVID-19 on the overall performance of the system. In terms of data, the analysis in this section utilizes two main datasets both of which provided by TAMU CAPS. The first dataset provides detailed patient-level record data while the second provides information regarding counselor scheduling. We provide a detailed description of the structure of each of these datasets below.

The patient-level record dataset provided by TAMU CAPS contains anonymized information about each session that has been scheduled at CAPS across the entire semester. Each row corresponds to a unique session for a patient and contains information such as the patient ID (a unique identifier for each patient), the session type (e.g., crisis, ongoing, or other), whether a cancellation or no-show occurred, the name of the assigned counselor, and the session date and time. This dataset is an integral component of our study since it provides detailed session-level information, and most of the input parameters for the simulation is obtained by analyzing this dataset. Before performing any analysis, however, the data was carefully reviewed in order to remove any potential anomalies such as technical issues related to data collection. For instance, there were several duplicate rows in the dataset which needed to be removed. There were also some instances where a patient had two ongoing sessions on the same day at different points of time. After lengthy discussions with TAMU CAPS, it was confirmed that these instances were due to technical errors, and were therefore removed from the dataset. Once the data was cleaned, we augmented it with some extra features to facilitate the analysis. For instance, we added a Boolean feature that identifies whether the session for each row corresponds to a crisis or a regular patient. Similarly, for each session/patient pair, we added a categorical feature that provides the session number that the patient is attending (i.e., first session, second session, etc.). The resulting dataset can then be analyzed to obtain the various input parameters of our DES model. The details of obtaining these parameters and their usage in the model are discussed in Section \ref{sec: IA}.

The counselor schedule dataset contains information about the number of sessions allocated to each service type by each of CAPS' counselors. This dataset contains a total of $36$ rows, one for each unique counselor at TAMU CAPS. Each column of this dataset provides information on the total time per week that the counselors are expected to allot for each unique service. Recall that other than direct counseling of patients via first-time, ongoing, or crisis sessions, counselors also need to allocate time for other services such as organizing workshops, intern supervision, student documentation, etc. For the purpose of this study, these time slots are allotted to the ``other'' service type as discussed in Section \ref{sec:scheduling}. It is important to note that the information provided in this dataset does not characterize a unique schedule topology. However, it provides an idea of the proportion of time that each counselor spends for each service category across the entire semester. In order to get a realistic representation of the schedule topology based on this dataset, we use the proportions of each service category in the schedule to generate random realizations of schedule topologies. In our experiment in Sections \ref{sec:validation} and \ref{sec:experiments}, we perform several replications of the simulation, each with a different random topology realization, and calculate the average estimates of the KPIs across the replications for a specific proportion distribution. Such an analysis is desirable because implementing a specific topology may be challenging. Instead, we aim to provide recommendations on the proportion of time that each counselor should ideally dedicate for each service type. This can be used as a master plan by CAPS and can potentially act as a guideline for the counselors to plan their schedules for the semester.

\subsection{Input Analysis} \label{sec: IA}

Input analysis is an important component of discrete-event simulation models as the simulation output heavily depends on the input. If a model's input parameters (e.g., arrival rates, no-show and cancellations probabilities, etc.) do not accurately represent real-life behaviors, then the output of a simulation could be unreliable even if the simulation logic is accurate. Therefore, the main purpose of input analysis is to theorize probability distributions that correctly describe the real-life system parameters. In this section, we describe the process of obtaining these input parameters from the previously mentioned patient-level record data.

\begin{figure}[t]
\centering
\caption{Number of new patient arrivals per week at TAMU CAPS in Fall $2019$.}
\includegraphics[trim=2.1cm 8cm 0.2cm 8cm, clip, width=.6\textwidth]{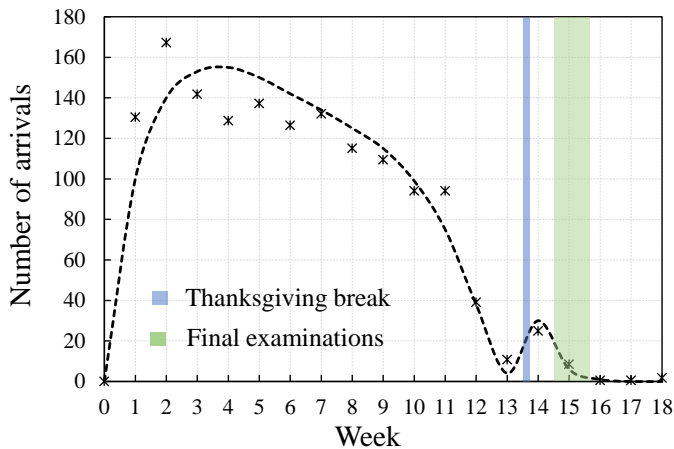}\label{fig:ARCaps}
\vspace{-0.5cm}
\end{figure}

A fundamental component of the simulation is the arrival rate of patients as it will play a vital role in describing the dynamics of the simulation model. To estimate the arrival rate, we use data on the total number of new patient arrivals on a per week basis. This data is summarized in Figure  \ref{fig:ARCaps} which plots the number of new patient arrivals as a function of time. The data clearly reveals an interesting pattern: As the semester starts, a rush of requests is observed. This demand gradually reduces with a sudden drop right before Thanksgiving break (shaded blue region). The demand then picks up right before  final exams (shaded green region). This pattern is consistently observed across semesters (e.g, for the spring semester a similar trend is observed however the drop in demand is observed right before Spring Break). Recall that part of the objective of this analysis is to take advantage of such predictable cyclical demand patterns. To obtain the per hour arrival rates from this data, we divide the total number of arrivals for each week with the total number of work hours per week (set to $40$). As common in the literature, we assume that arrivals follow a non-stationary poison process with arrival rates obtained from Figure \ref{fig:ARCaps}.

It is important to note that the above arrival rate takes into account arrivals of both regular and crisis patients. Consequently, determining whether the arriving patient is a crisis or a regular patient is a core element in modeling the arrival process and the simulation. This is because upon arrival, the subsequent processes that patients undergo are dependent on whether they are a regular or a crisis patient. In order to model the differences in arrival rates for regular and crisis patients, we adopt a Poisson split. This is a commonly used technique that allows for splitting a Poisson process into two separate Poisson processes, each with a different arrival rate. This split is based on the proportions of different entities that are expected to arrive in the simulation. For this case study, we use the proportion of crisis and regular patients as the proportions for the Poisson split. Using the patient-level record data, the ratio of crisis patients, obtained by dividing the number of patients who received at least one crisis session by the total number of patients, is equal to $9.6\%$. As a result, the ratio of regular patients  is equal to $90.4\%$. Thus, whenever an arrival occurs, there is a $9.6\%$ ($90.4\%$) chance that the patient is a crisis (regular) patient.

\begin{figure}[!t]
  \centering
  \caption{Distribution of total number of sessions for regular and crisis patients.}
  \subfloat[Regular patients]{\includegraphics[trim=0.2cm 0.2cm 0.2cm 0.2cm, clip, width=0.5\textwidth]{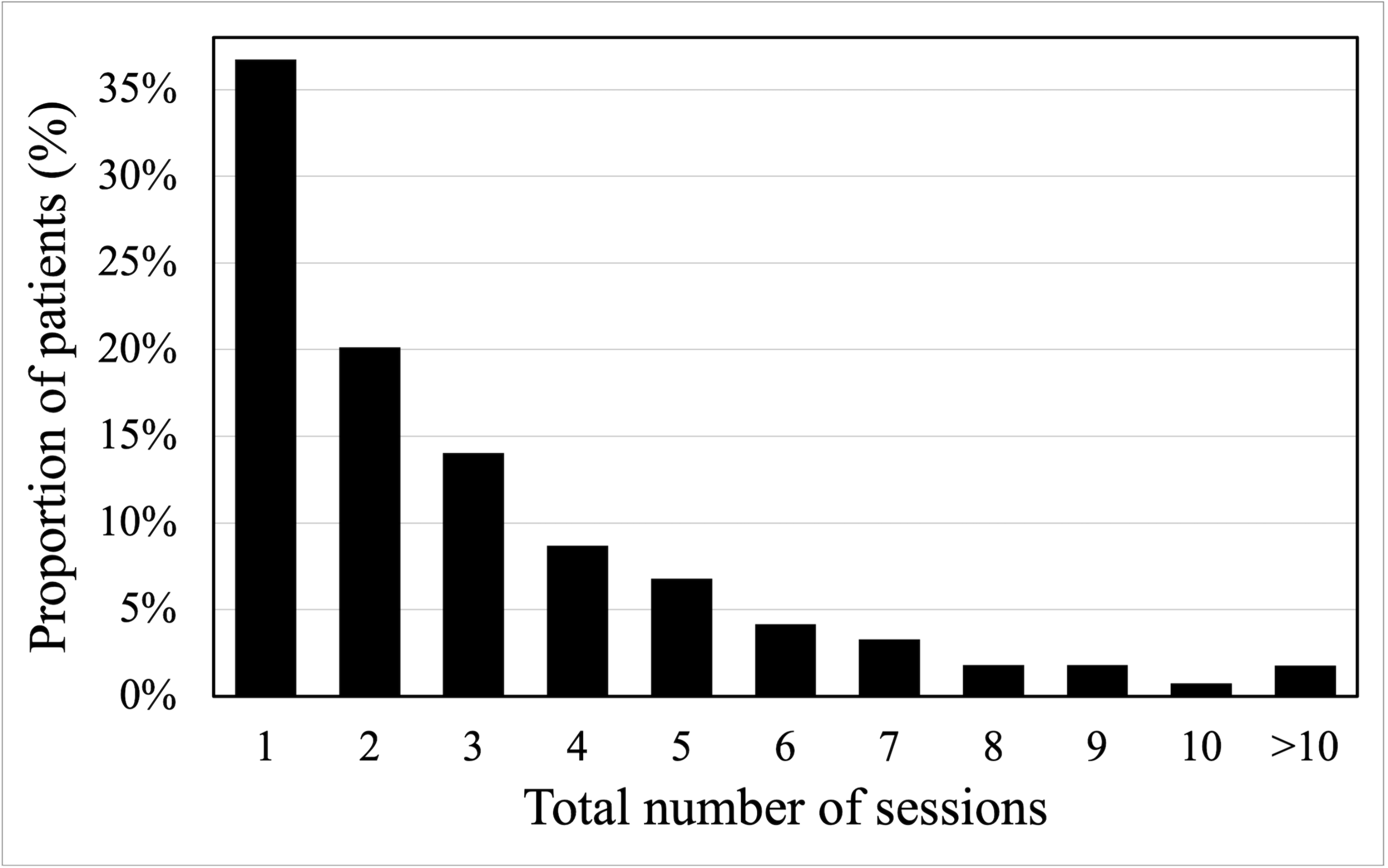}\label{fig:regsesdistregclnts}}
  \hfill
  \subfloat[Crisis patients]{\includegraphics[trim=0.2cm 0.2cm 0.2cm 0.2cm, clip, width=0.5\textwidth]{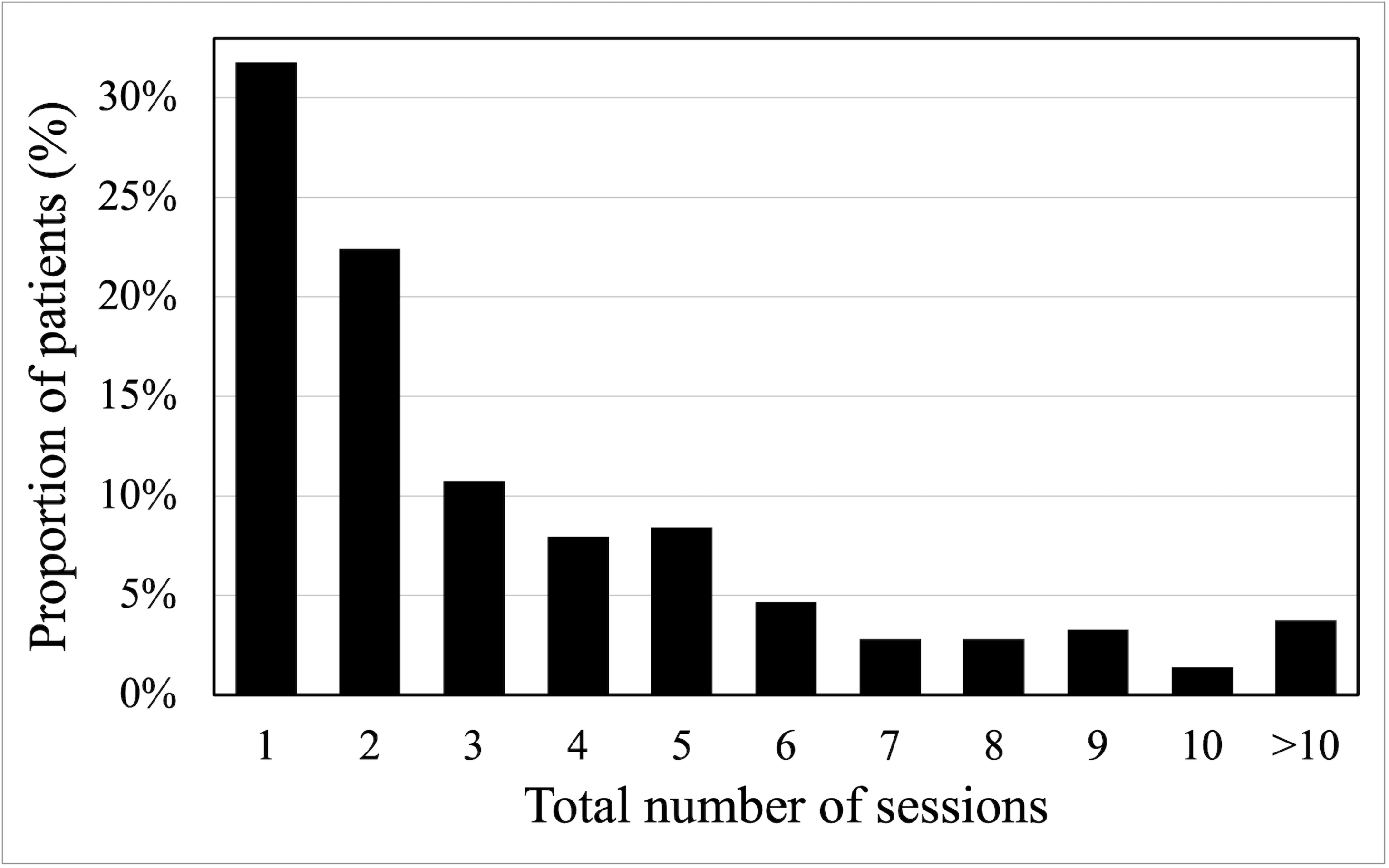}\label{fig:regsesdistcrsclnts}}
  \vspace{-0.5cm}
\end{figure}

Upon arrival, several important initial features are allotted to each patient. One of the most important features is the number of sessions that the patient has to attend. Of course, different patients can require a different total number of sessions, which is typically determined by their counselor. In order to allow for this variability in the total number of sessions, we use a probability distribution based on information obtained from the patient-level record dataset. The distributions of the total number of sessions for regular and crisis patients are shown in Figures \ref{fig:regsesdistregclnts} and \ref{fig:regsesdistcrsclnts}, respectively. Although somewhat similar, the two distributions have some unique characteristics which highlights the difference in treatment patterns between regular and crisis patients. For instance, it can be observed that crisis patients are more likely to utilize a higher number of sessions, which emphasizes their need for CAPS' services for a relatively longer period. Another important distinction is that regular patients have a higher probability of requiring just one session, while crisis patients have a higher tendency of requiring more sessions. This can be attributed to the fact that crisis patients arriving into the system during an emergency crisis attack typically schedule a follow-up session with the counselor. Using these distributions, and depending on the arriving patient type, a random number of total sessions is generated for each arrival.

Another important initial feature that is specific to crisis patients is the crisis attack probability. Recall that crisis patients may arrive into the CAPS system during an emergency crisis attack. However, the patient-level record data also reveals that crisis attacks can occur at any point in between sessions. This aspect is modeled into the simulation using the crisis attack probability. At the end of a session, and once the next appointment has been decided upon by the counselor and the patient, crisis patients have a probability (characterized by the crisis attack probability) of experiencing a crisis attack before the next scheduled session. This probability is estimated from the patient-level record dataset filtered on only crisis patients. Then, for each crisis patient, a crisis attack probability is obtained by dividing the number of crisis sessions that occurred between two sessions by the total number of gaps between every two sessions. The overall crisis attack probability is then calculated by taking the average of all individual probabilities, and is found to be $14.6\%$. That is, after the end of a session a crisis patient has a $14.6\%$ probability of experiencing a crisis attack before their next scheduled session. 
    
\begin{table}[t]
\centering
\renewcommand{\arraystretch}{1.5}
\begin{tabular}{l M{3cm} M{3cm}}
& Regular patients& Crisis patients\\\hline
No-show probability& 9.7\% & 8.9\% \\ 
No-show exit probability &58.7\%&39.5\%\\ 
Cancellation probability    &18.6\%&15.7\%\\ 
Cancellation exit probability &38.4\%&27.1\%\\ \hline
\end{tabular}
\caption{No-show and cancellation-related probabilities for regular and crisis patients.}
\label{tab:NS_CanProb}
\end{table}

Once the initial features have been allotted, patients can book their first-time session and then continue their treatment cycle with the assigned counselor until they meet the total number of sessions that was assigned to them. However, during their course of sessions at CAPS, patients may cancel or not show up for their scheduled session. As mentioned in Section \ref{sec: Sim}, the no-show and cancellation probabilities are two key features assigned to each patient entity upon their arrival into the system. In the same context, there are two other probabilities assigned to each patient: The no-show exit and the cancellation exit probabilities. Our discussions with CAPS, as well as an analysis of the patient-level record data, revealed that in the event of a no-show or a cancellation there are instances where patients do not reschedule their appointment and simply discontinue their visits to CAPS. For instance, the data reveals that roughly $60\%$ of the time regular patients discontinue their visits to CAPS after a no-show, while crisis patients only do so about $40\%$ of the time. This trend is captured in our simulation using two additional features: The no-show exit and cancellation exit probabilities. Once a no-show occurs, the probability that the patient exits the system without continuing their remaining sessions is represented by the no-show exit probability. The cancellation exist probability, on the other hand, represents the probability of the patient exiting the system following a cancellation. These probabilities can vary depending on the patient type, i.e., regular or crisis. To obtain the no-show (or cancellation) probabilities we divide the number of no-shows (or cancellations) by the total number of sessions. Similarly, the no-show exit (or cancellation exit) probabilities are obtained by dividing the number of cases where the patient had a no-show (or cancellation) as their last session by the total number of no-shows (or cancellations). This process is performed separately for regular and crisis patients. All four of the aforementioned probabilities are shown for each patient type in Table \ref{tab:NS_CanProb}. Observe that the no-show and cancellation probabilities are lower for crisis patients in comparison to regular patients, which highlights that crisis patients have a higher inclination to attend their counseling sessions. It is also observed that the exit probabilities are notably lower for crisis patients, which alludes to the fact that crisis patients have a higher tendency to complete their full course of sessions at CAPS in comparison to regular patients.

\begin{figure}[!t]
\centering
\caption{Empirical distribution of the time between sessions.}
\includegraphics[trim=0.1cm 0.2cm 0.1cm 0.1cm, clip, width=0.75\textwidth]{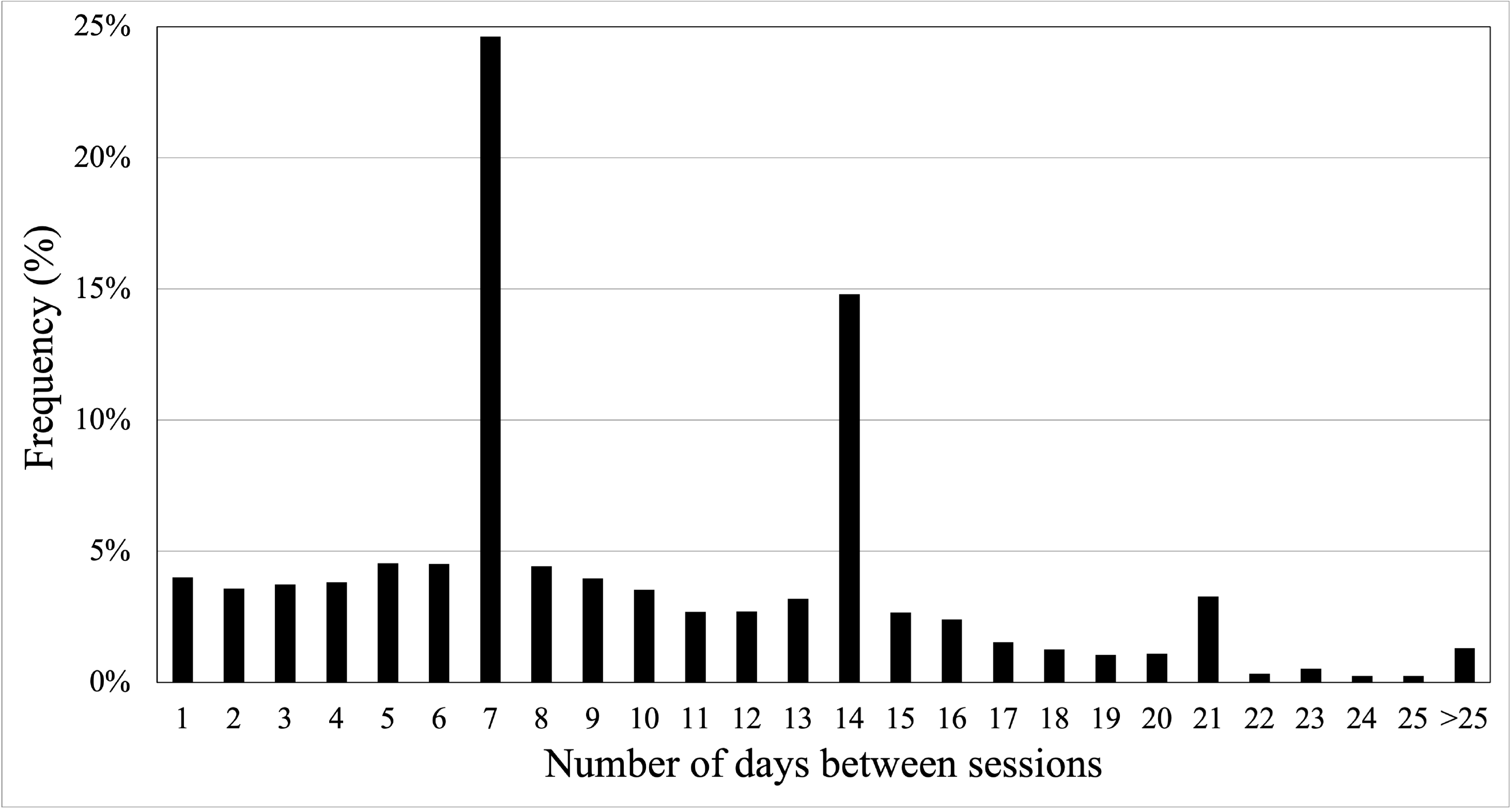}\label{fig:timebtwsessions}
\vspace{-0.4cm}
\end{figure}

At the end of each session, the counselor and the patient need to decide on a date and time for their next meeting (this excludes situations in which the patient is done with their total number of sessions). The time between sessions is typically a function of the counselors expert opinion on session frequency (which might be case specific), as well as schedule availability. In fact, an analysis of the number of days between sessions performed on the patient-level record dataset reveals some interesting features (see Figure \ref{fig:timebtwsessions} which provides the empirical distribution of the time between sessions) that can be explained on the basis of these two factors. Observe that the figure clearly depicts three distinct spikes corresponding to $7$, $14$, and $21$ days, which highlights the fact that counselors conventionally prefer meeting on a weekly, bi-weekly, or tri-weekly basis. We suspect that this is based the counselor's expert opinion on when then next session should ideally occur. Another key observation is that the proportion of cases of $7$ days is much more common than cases for $14$ or $21$ days. This indicates that counselors often see the need to schedule weekly appointments.     Finally, note that the lower proportions around the peaks can be explained by the second factor, i.e., the inability (due to slot availability) to find a time on exactly $7$, $14$, or $21$ days after their session. As a result, based on schedule availability, the two parties might agree on a day in and around the initially estimated preferred date. This is an important feature of the operations at CAPS that needs to be intricately modeled into the simulation. To achieve this, while determining the date for the next session, we first pick an initial preferred date based on the probabilities of the three peaks. If that exact date is unavailable (i.e., no ongoing sessions slots are available), we start searching for available slots around that day until one is found. Doing so incorporates both of the aforementioned factors when deciding the next patient appointment time.

\subsection{Model Validation}\label{sec:validation}
 
An important step before conducting the simulation experiments is to validate the model. That is, to determine whether the simulation is accurately representing the real-world performance. To achieve this, we run the simulation model using the Fall $2019$ datasets and compare the system performance measured from the output of the simulation model to the reported Fall $2019$ performance of the system. We mainly focus on the access time as it is the main performance metric reported by CAPS. To run this experiment, we first need to define the current schedule topology being implemented in TAMU CAPS. Unfortunately, as discussed in Section \ref{sec:datades}, the counselor schedule dataset provided by CAPS does not fully characterize a schedule topology. Instead, it provides information about the proportion of time that each counselor spends for each service type across the entire semester. In this study, in an effort to capture the wide spectrum of topologies that adhere to these proportions, we utilize the information provide by the counselor schedule dataset to generate representative random topology realizations. For each of these realizations, the simulation is executed and the expected access time is stored.

\begin{figure}[!t]
  \centering
  \caption{Boxplot for expected access time under the currently implemented TAMU CAPS operating parameters.}
  \includegraphics[trim=0.1cm 3.5cm 0.1cm 3.8cm, clip, width=0.7\textwidth]{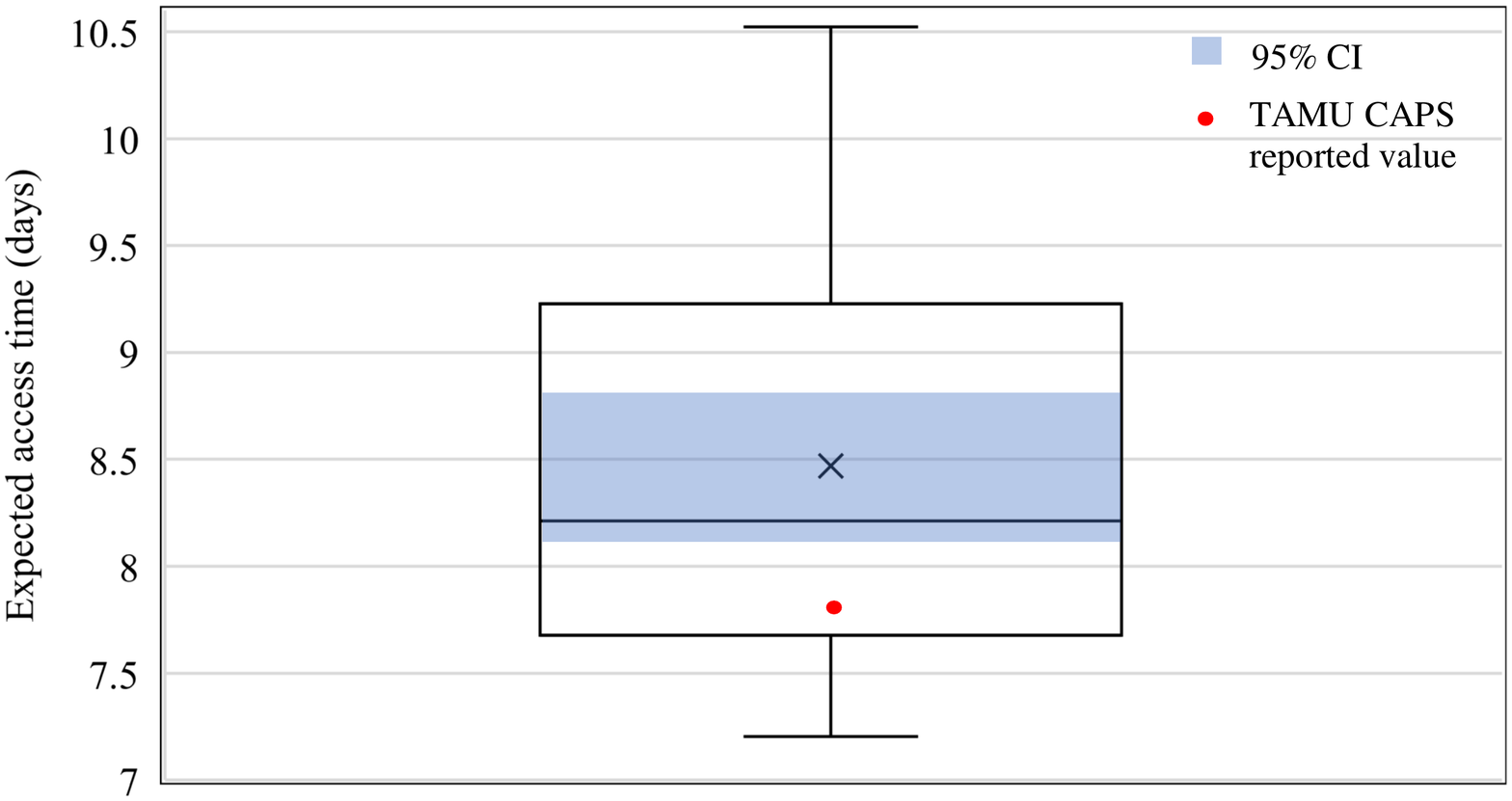}\label{fig:ratbox}
\end{figure}

To obtain the current proportions from the counselor schedule dataset, we sum the total number of hours dedicated to each service type for all of the counselors across the semester. Then, assuming a $40$ hour work week and a time span of $18$ weeks, we divide the sum of time allocated for each service type by the total work time across the semester. Doing so reveals that, on average, counselors spend $8\%$ of their time on first-time sessions, $4\%$ on crisis sessions, $25\%$ on ongoing sessions, and the remaining $63\%$ of the time is dedicated for ``other'' services. To generate a topology realization that adheres to these proportions, we execute a pre-processing procedure at the very beginning of the simulation on the external excel file discussed in Section \ref{sec: Sim}. This procedure enumerates over all rows of the file and assigns, for each counselor, a random service type ID that is generated based on the distribution obtained from the counselor schedule dataset. Specifically, each time slot has an $8\%$ probability of being assigned a first-time service type, $25\%$ probability of being assigned an ongoing service type, $4\%$ probability of being assigned a crisis slot, and a $63\%$ probability of being assigned a ``other'' service type. The resulting schedule topology ensures that, on average, the proportion of time committed to each service type by each counselor matches that of the counselor schedule dataset. Once a schedule topology is realized, the simulation is executed and the expected access time is recorded. This entire procedure is repeated a number of times to generate a distribution of expected access times. Note that while the overall proportions across iterations are preserved, the specific schedule topology can vary. Such an experiment can provide valuable information about the variability of the system performance for a fixed proportion distribution.

We run the aforementioned procedure for a total of 32 replications. Note that each replication represents a simulation run across the entire semester for a randomly generated schedule topology. Figure \ref{fig:ratbox} shows the box and whisker plot for the expected access time across the $32$ replications of our experiment.
The Figure reveals a number of interesting behaviors: First, for a fixed proportion distribution, the expected access time can vary quite substantially with values ranging from $7.2$ days to as high as $10.5$ days. This indicates that, even when the proportion distribution is fixed, the schedule topology can have a significant impact on system performance and a carefully designed schedule topology can improve system performance by up to $46\%$. Second, the average expected access time is obtained to be $8.5$ days (with a half width of $0.4$ days), which closely aligns with the actual reported access time of $7.8$ days. This observation is important as it indicates that the simulation is accurately describing the real-world performance of the system. Notice, however, that the reported $7.8$ days does not coincide with the simulation confidence interval (represented by the shaded blue region). 
That is, the simulation is slightly overestimating the expected access time. This behavior is somewhat expected as the simulation experiment was executed for random schedule topologies. In practice, however, CAPS directors put effort in designing good performing topologies which are expected to outperform randomly generated topologies. It is interesting to note that TAMU CAPS is doing a good job in identifying good performing schedule topologies with only $31\%$ of the randomly generated topologies outperforming the current implementation. However, based on our experiments, it is possible to further improve the performance of the current implementation by up to $8\%$.

\subsection{Simulation Experiments and Sensitivity Analysis}\label{sec:experiments}

Having ensured that the simulation model is accurately depicting the real-world system perform, in this section we perform a series of numerical experiments. As discussed at the beginning of Section  \ref{sec: casestudy} we conduct three experiments to investigate the impact of: (i) varying the proportion of external referrals, (ii) imposing a maximum session limit, and (iii) the structure of the schedule topology. These three experiments are outlined below.

\begin{table}[t]
        \centering
        \renewcommand{\arraystretch}{1.3}
        \begin{tabular}{M{5.5cm} | M{4.5cm} M{4.5cm}}
            Proportion of external referrals & Access time in days (HW) & Crisis time in hours (HW)\\ \hline
             2.8\%&8.66 (0.32)&1.64 (0.11)\\
            4.2\%&8.48 (0.39)&1.65 (0.17)\\
            5.6\%&8.47 (0.37)&1.81 (0.20)\\
            7.0\%&8.81 (0.35)&1.71 (0.18)\\
            8.4\%&8.65 (0.30)&1.70 (0.16)\\
            9.8\%&8.60 (0.39)&1.72 (0.21)\\
            11.2\%&8.73 (0.38)&1.67 (0.14)\\\hline
        \end{tabular}
        \caption{Expected access and crisis times for varying proportion of external referrals.}
        \label{tab:refprop}
    \end{table}
 
In the first experiment, we vary the proportion of external referrals from CAPS and observe its impact on the access time and the crisis time. As mentioned in Section \ref{sec:introduction}, in order to meet a rise in demand, various university counseling centers, including TAMU CAPS, often refer students to external off-campus mental health providers, especially if they require a higher level of specialization. Typically, it is only after a few sessions that the counselor is in a position to judge whether the patient needs to be externally referred. Once that decision has been made and conveyed to the patient, they still attend a few more sessions with their regular counselor at CAPS in order to ensure a smooth transition to their new provider. Based on an analysis of the Fall $2019$ patient-level record data, the proportion of patients that were referred to external providers was identified to be $5.6\%$. Other relevant information, such as the average number of sessions attended prior to a referral as well as the average number of transition sessions after a referral, is also determined from the same dataset and incorporated into the simulation model. The natural question in this regard is whether changing the proportion of external referrals influences the two KPIs of interest, i.e., the access time and the crisis time. We run a series of experiments by varying the external referral proportion from half of its current value, i.e., $2.8\%$ to double, i.e., $11.2\%$. Other input parameters, such as the proportion distribution of the schedule topology and the maximum session limit is maintained at the current operating values for consistency. The results of this numerical experiment are summarized in Table \ref{tab:refprop}. The results reveal that neither the expected access time nor the expected crisis time are impacted by changes in the proportion of external referrals.  
\begin{figure}[t!]
  \centering
  \caption{Expected access time (a) and crisis time (b) as a function of the proportion of external referrals.}
  \subfloat[Expected access time]{\includegraphics[trim=0.2cm 3cm 0.2cm 3cm, clip, width=0.5\textwidth]{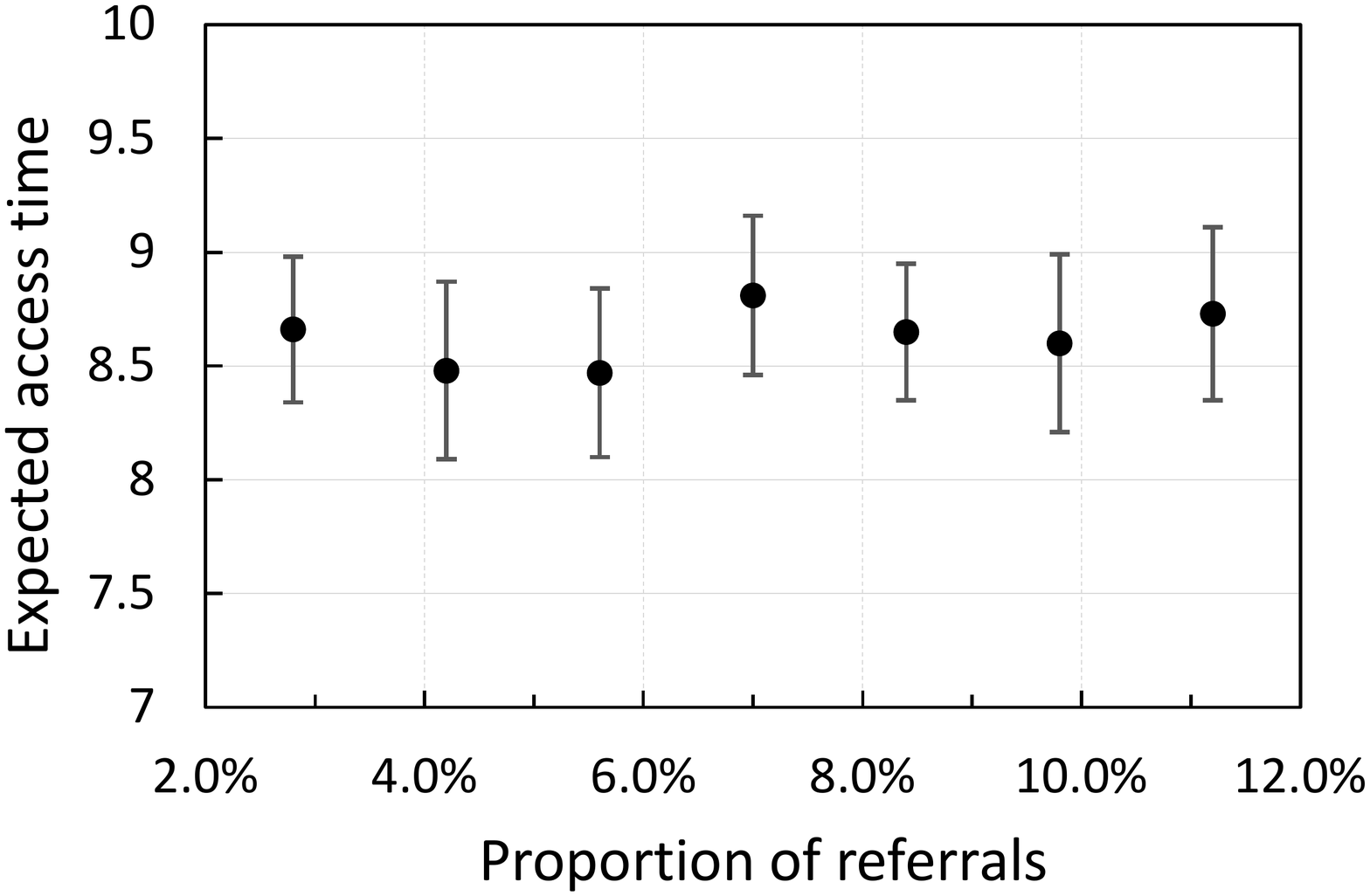}\label{fig:referralAccess}}
  \hfill
  \subfloat[Expected crisis time]{\includegraphics[trim=0.75cm 8.2cm 0.75cm 8.2cm, clip, width=0.5\textwidth]{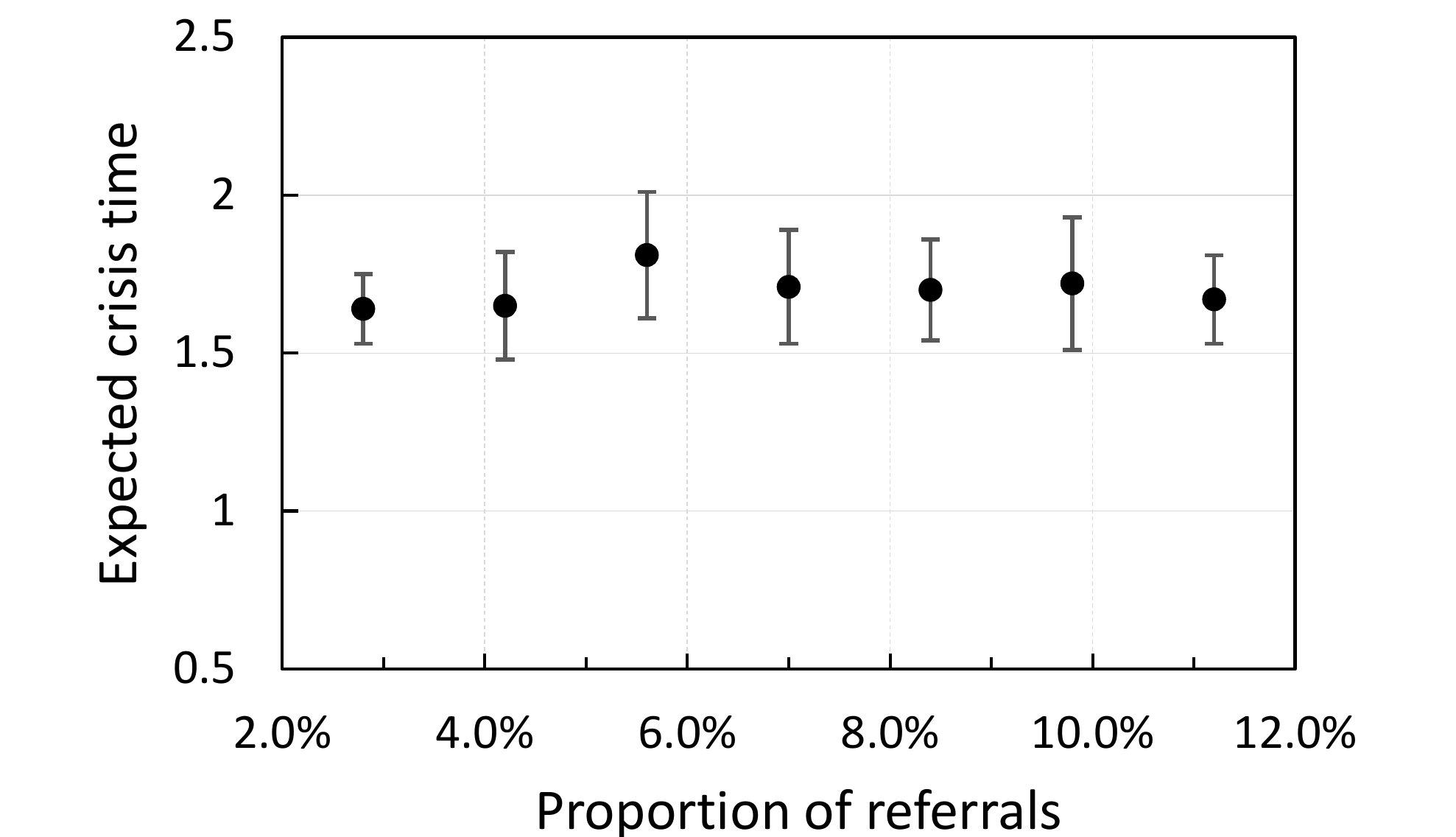}\label{referralCrisis:maxCapCrisis}}
  \label{fig:referral}
  \vspace{-0.8cm}
\end{figure}
This is more clearly depicted in Figure \ref{fig:referral} which plots the expected value of both KPIs (along with the confidence intervals) as function of the proportion of external referrals. This is apparently counter-intuitive since one would expect that external referrals would improve access to new patients. However, note that according to the current TAMU CAPS policy, students are only referred to external providers once they are done with a certain number of sessions. Thus, although external referrals may allow for extra available ongoing slots, they still exhaust valuable first-time slots. As a consequence, it does not impact the access time since even if we increase the referral proportion from $2.8\%$ to $11.2\%$, it still ends up utilizing the same number of first-time slots on average. A similar observation can be made with regard to the crisis time. Such results demonstrate how solely changing the proportion of external referrals without modifying the schedule topology will not have the desired impact on the overall performance of the system.

\begin{figure}[!t]
  \centering
  \caption{Expected access time (a) and crisis time (b) as a function of the maximum session limit.}
  \subfloat[Expected access time]{\includegraphics[trim=0.2cm 3cm 0.2cm 3cm, clip, width=0.5\textwidth]{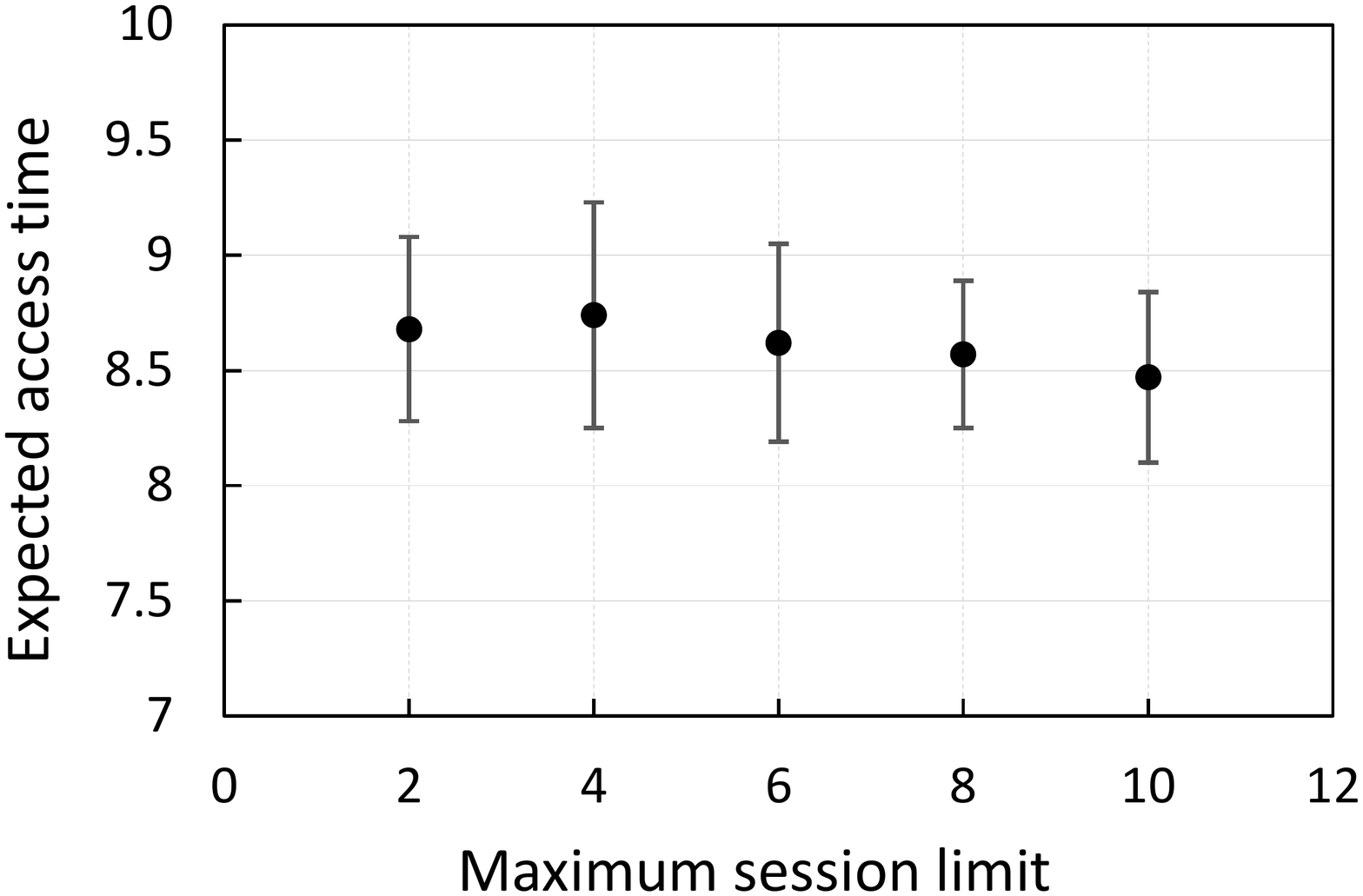}\label{fig:maxCapAccess}}
  \hfill
  \subfloat[Expected crisis time]{\includegraphics[trim=0.75cm 8.2cm 0.75cm 8.2cm, clip, width=0.5\textwidth]{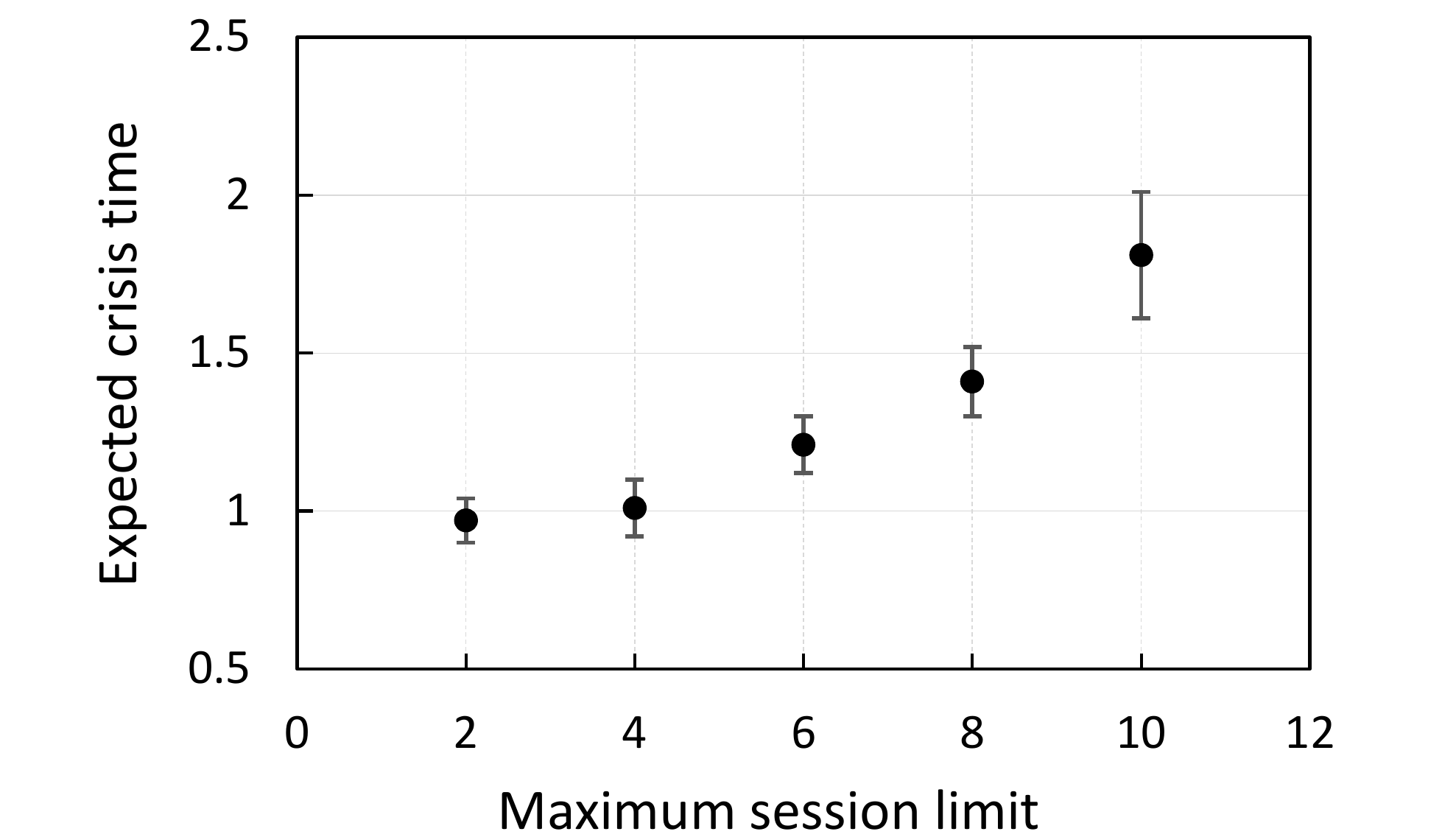}\label{fig:maxCapCrisis}}
  \vspace{-0.5cm}
\end{figure}

For the next set of experiments we perform a sensitivity analysis on the maximum number of sessions allowed per patient, which is a concept that was introduced in Section \ref{sec:introduction}. Such a policy is introduced in various CAPS systems to increase access to newer patients. In our model, this policy is implemented in the following manner: We generate a random number of total sessions from the distribution described in Section \ref{sec: IA}, and then select the minimum value among the generated number and the maximum session limit as the number of sessions to be allotted to the patient. Next, for our experimental setup, we consider a range of values for the maximum session limit ($2$ to $10$), and run a total of $16$ replications for each case to observe its impact on the average values of the two KPIs. The other parameters, such as the topology proportion distribution and the proportion of external referrals, are set to the current TAMU CAPS operating values. 
\begin{table}[t!]
\centering
\renewcommand{\arraystretch}{1.5}
\begin{tabular}{M{4.5cm} | M{4.5cm} M{4.5cm}}
Maximum session limit & Expected access time in days (HW) & Expected crisis time in hours (HW)\\\hline
2 &8.68 (0.40)& 0.97 (0.07)\\
4    &8.74 (0.49) & 1.01 (0.09)\\
6 & 8.62 (0.43)  &  1.21 (0.09)\\ 
8 & 8.57 (0.32) & 1.41 (0.11)\\ 
10& 8.47 (0.37) & 1.81 (0.20)\\ 
\hline
\end{tabular}
\caption{Expected access and crisis times for varying maximum session limits.}
\label{tab:Maxsescap}
\end{table}
The results for this experiment are summarized in Table \ref{tab:Maxsescap}, which reports the mean values of our two KPIs as well as the half-width (HW) of a $95\%$ confidence interval in parenthesis. Observe that changing the maximum limit does not lead to any statistically significant changes in the expected access time. This is more clearly demonstrated in Figure \ref{fig:maxCapAccess}. Although this can seem anomalous, some insight into the operating scheme of TAMU CAPS makes the observation quite intuitive. This is because the access time is the time taken from the patient's request for a first-time session to the appointment time. The primary factor that impacts this time is the proportion of first-time slots available in the schedule topology. Since the schedule topology remains the same across the cases on average, merely changing the maximum session limit does not influence the access time for patients. More interestingly, the crisis time is increasing with the maximum limit. Again, this is more clearly demonstrated in Figure \ref{fig:maxCapCrisis}. This can be explained by the fact that as patients tend to attend a higher number of sessions, it increases the probability of having crisis attacks. Clients attending more sessions (which is a consequence of a higher maximum limit) will inevitably have a higher probability of experiencing a crisis attack (see Section \ref{sec: IA}). Since the proportion distribution is preserved in this experiment, then the total number of crisis slots in the schedule remains the same. This increase in demand for fixed number of crisis slots ultimately leads to higher crisis times.

The first two experiments revealed that, for a fixed proportion distribution, the expected access time is not impacted when varying the proportion of external referrals and the maximum session limit. To observe the impact of the proportion distribution on the two KPIs, we run a two-dimensional sensitivity analysis on the proportions of first-time slots ($p_f$) and  crisis slots ($p_c$). We vary these values while fixing the proportion of slots dedicated to the ``other'' category. Consequently, the change in $p_f$ and $p_c$ will ultimately impact the proportion of ongoing slots (to preserve a total proportion of $1$). Following the approach discussed in Section \ref{sec:validation}, we generate a random topology for a given pair $(p_f,p_c)$ and run the simulation for the entire semester. This process is repeated for a total of 25 replications. We conduct this analysis for a range of $p_f$ values in $\{5\%,7\%,9\%,11\%,13\%,15\%,17\%\}$ and $p_c$ values in $\{1\%,3\%,5\%,7\%,9\%\}$. The results are summarized in Table \ref{tab:RATvsCS} which reports the expected access and crisis times (and the half width in parentheses)  for all scenarios. The results reveal several interesting behaviors, discussed below.

\begin{figure}[t!]
  \centering
  \caption{Expected access time when $p_c=1\%$ as a function of $p_f$ (a) and expected crisis time when $p_f=5\%$ as a function of $p_c$ (b).}
  \subfloat[Expected access time]{\includegraphics[trim=0.2cm 3cm 0.2cm 3cm, clip, width=0.5\textwidth]{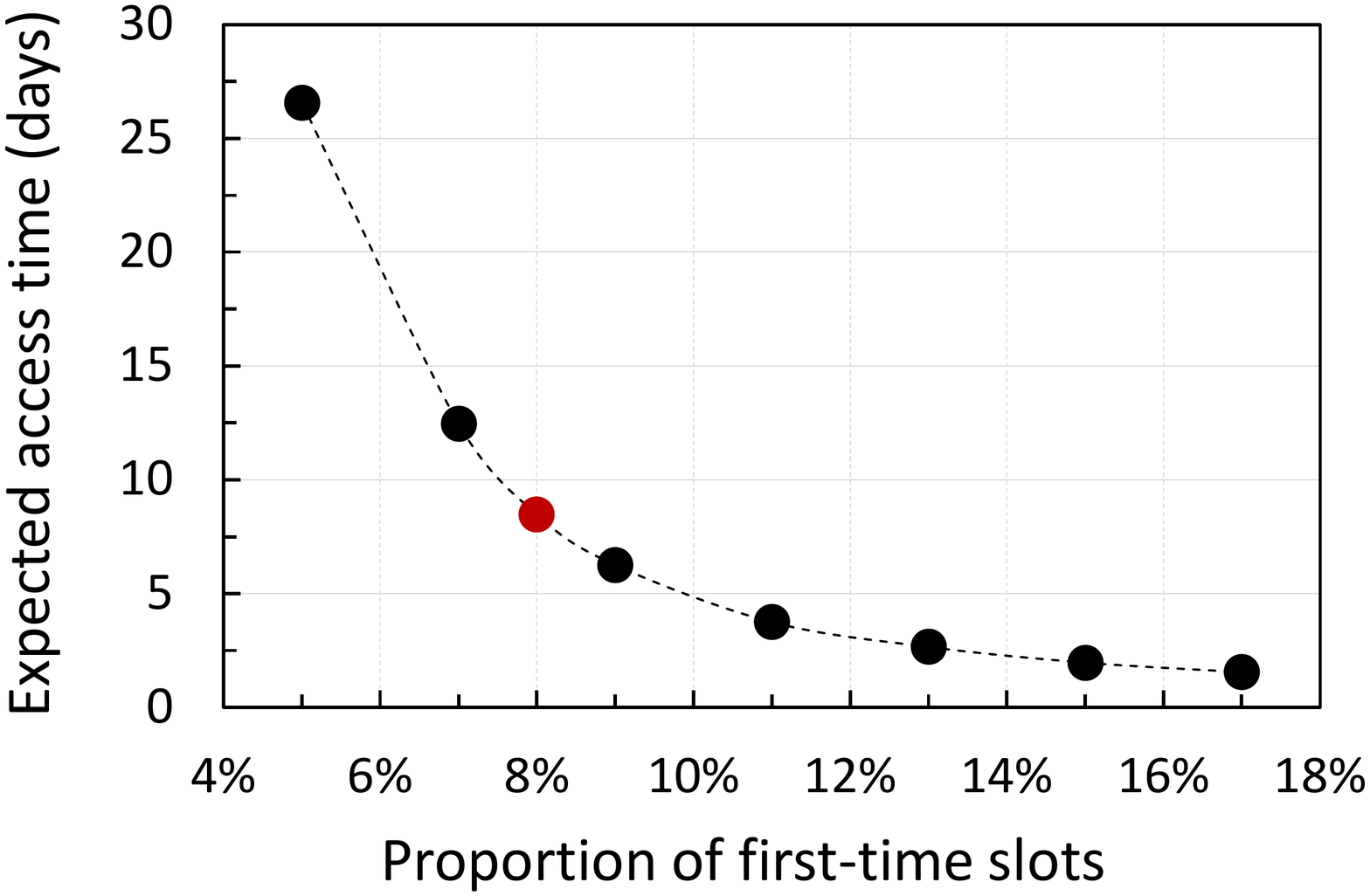}\label{fig:accessFirst}}
  \hfill
  \subfloat[Expected crisis time]{\includegraphics[trim=0.2cm 3cm 0.2cm 3cm, clip, width=0.5\textwidth]{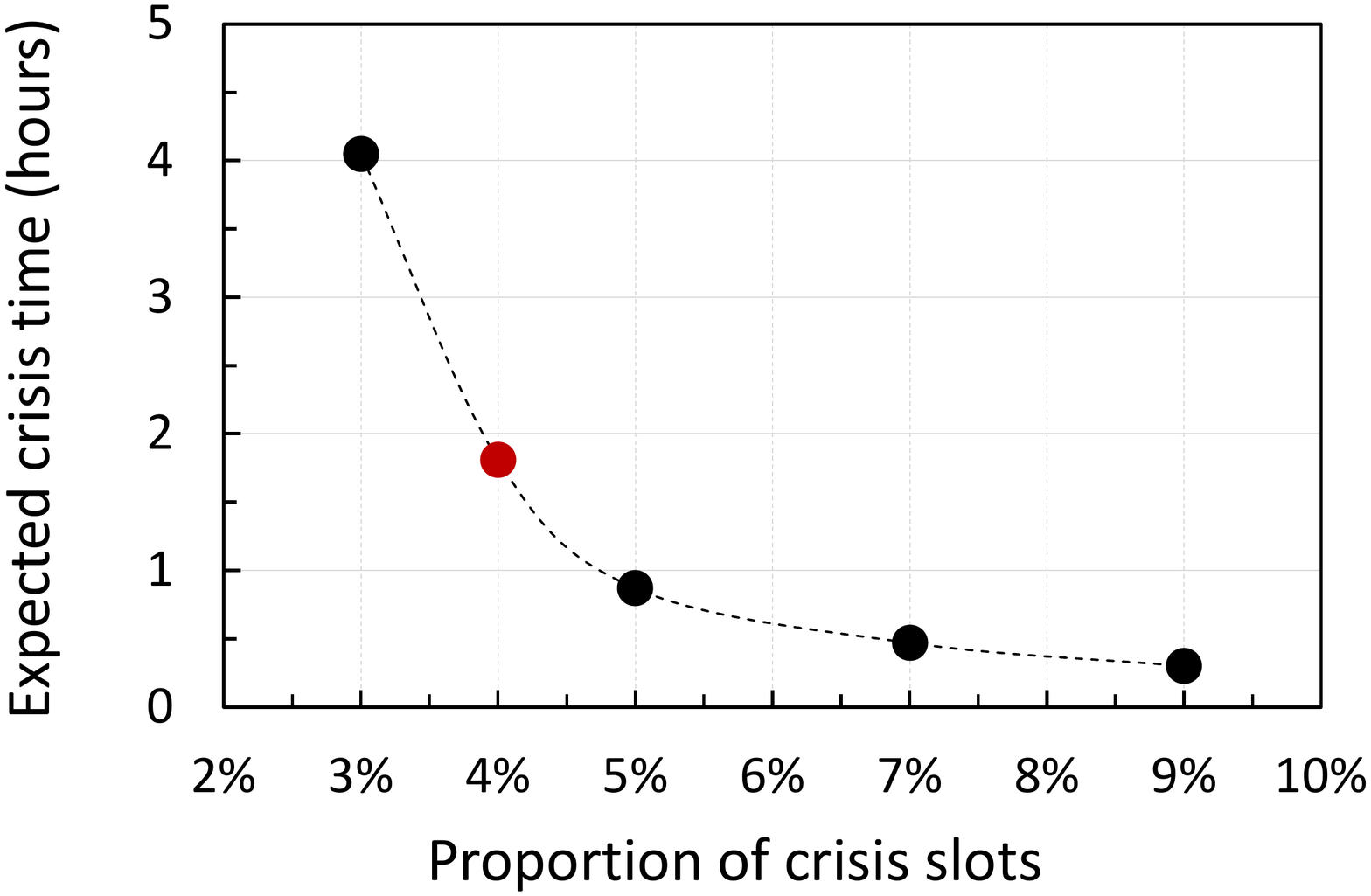}\label{fig:crisisCrisis}}
  \vspace{-0.5cm}
\end{figure}

Looking at the expected access time, the results show that it is heavily dependent on $p_f$ with higher proportions leading to lower expected access times. Note that even small variations can lead to large differences. For example, reducing $p_f$ from $8\%$ (current implementation) to $7\%$ increases the expected access time by a substantial $50\%$ (on average). Increasing $p_f$ to $9\%$, on the other hand, reduces the expected access time by an average of $26\%$. 
The sensitivity of the expected access time to $p_f$, however, exhibits diminishing returns. To better see this, Figure \ref{fig:accessFirst} plots the expected access time when $p_c=1\%$ as a function of $p_f$. Such an observation is important because increasing the value of $p_f$ beyond a certain point may not be worthwhile since it will only lead to a marginal improvement in the system performance. The results also reveal that the expected access time is not impacted by $p_c$. This is somewhat expected  since varying the proportion of the topology dedicated to crisis should not impact the access time. Such results can help CAPS directors determine topology proportions that meet their ideal performance. For example, if CAPS aims to have an expected access time of just 3 days, then our results can be used to determine a $p_f$ value that will attain this performance ($p_f\approx 12.5\%$).

Shifting the focus to the expected crisis time, Table \ref{tab:RATvsCS} reveals similar trends. In particular, the expected crisis time is decreasing with $p_c$ with the greatest reduction observed for lower $p_c$ values. This diminishing returns feature is more clearly depicted in Figure \ref{fig:crisisCrisis}. For example, reducing $p_c$ from $4\%$ (current implementation) to $3\%$ increase the expected crises time by an average of 130\% while increasing $p_c$ to $5\%$ improves the expected crisis time by an average of $50\%$. Also, the results reveal that the expected crisis time does not seem to be correlated with $p_f$. However, when $p_c$ is substantially low, a slight positive correlation is observed with $p_f$. For example, when $p_c=1\%$ a $p_f$ of $5\%$ leads to an expected crisis time of $61.24$ hours while a $p_f$ of $17\%$ leads to an expected crisis time of $71.06$ hours. We suspect that this positive correlation is attributed to the fact that a higher $p_f$ value results in lower proportions for ongoing slots. This, in turn, would imply that the time between session would be prolonged (due to the lack of available slots) and in such a case the probability of the patient experiencing a crisis attack would increase. This increase in demand would then result in higher expected crisis times. However, this effect is only observed when the number of crisis slots is extremely low which is often not the case in practice. Again, these results can be used to identify values of $p_c$ that result in desirable performance. For example, if the target expected waiting time is 15 minutes, then, based on our analysis, a $p_c$ value of approximately $9.3\%$ would be needed to attain this performance.

 \begin{table}[t!]
\begin{threeparttable}
\small
\caption{Comparison of expected access time (EAT) and expected crisis time (ECT) with varying proportion of first-time ($p_f$) and crisis ($p_c$) sessions. Each cell reports the mean KPI value (in days for EAT and hours for ECT) and the half width for the $95\%$ confidence interval (in parenthesis).}
\label{tab:RATvsCS}  
    \begin{tabular}{M{11mm}| M{10mm} M{19mm} M{19mm} M{19mm} M{19mm} M{19mm} }
    \diagbox{$p_f$}{$p_c$} &&1\%&3\%&5\%&7\%&9\%\\\toprule  
    \multirow{2}{*}{5\%}
    &\textbf{EAT}&26.56 (0.67)&26.13 (0.76)&26.81 (0.81)&26.25 (0.80)&26.31 (0.88)\\

    &\textbf{ECT}&61.24 (3.75)&4.05 (0.53)&0.87 (0.10)&0.47 (0.05)&0.30 (0.03)\\
    \cline{2-7}
    \hline\\
    \multirow{2}{*}{7\%}
    &\textbf{EAT}&12.46 (0.49)&12.75 (0.57)&12.82 (0.67)&12.93 (0.64)&12.92 (0.75)\\

    &\textbf{ECT}&64.16 (3.76)&4.28 (0.82)&0.89 (0.08)&0.45 (0.04)&0.27 (0.03)\\
    \cline{2-7}
    \hline\\
    \multirow{2}{*}{9\%}
    &\textbf{EAT}&6.25 (0.30)&6.30 (0.29)&6.27 (0.30)&6.2 (0.28)&6.29 (0.33)\\
    &\textbf{ECT}&66.09 (3.86)&5.05 (1.25)&0.93 (0.07)&0.46 (0.05)&0.26 (0.02)\\
    \cline{2-7}
    \hline\\
    \multirow{2}{*}{11\%}
    &\textbf{EAT}&3.74 (0.13)&3.72 (0.16)&3.79 (0.17)&3.7 (0.17)&3.71 (0.18)\\
    &\textbf{ECT}&67.82 (3.07)&4.74 (0.82)&0.95 (0.10)&0.44 (0.05)&0.27 (0.02)\\
    \cline{2-7}
    \hline\\
    \multirow{2}{*}{13\%}
    &\textbf{EAT}&2.65 (0.10)&2.56 (0.08)&2.59 (0.09)&2.56 (0.06)&2.52 (0.09)\\
    &\textbf{ECT}&70.17 (3.09)&3.75 (0.58)&0.94 (0.09)&0.45 (0.05)&0.29 (0.03)\\
    \cline{2-7}
    \hline\\
    \multirow{2}{*}{15\%}
    &\textbf{EAT}&1.96 (0.07)&1.95 (0.05)&1.97 (0.06)&1.93 (0.05)&1.97 (0.06)\\
    &\textbf{ECT}&69.78 (3.51)&3.84 (0.85)&0.90 (0.08)&0.42 (0.04)&0.28 (0.03)\\
    \cline{2-7}
    \hline\\
    \multirow{2}{*}{17\%}
    &\textbf{EAT}&1.54 (0.04)&1.57 (0.05)&1.58 (0.05)&1.56 (0.05)&1.54 (0.04)\\
    &\textbf{ECT}&71.06 (3.44)&3.34 (0.58)&0.81 (0.07)&0.42 (0.05)&0.29 (0.03)\\
    \hline
    \end{tabular}
\end{threeparttable}
\end{table}

\section{Conclusions and Future Research Directions} \label{sec: conclusion}
In this paper, we construct a discrete-event simulation model that mimics the operational flow of college counseling centers. 
The considered model is general and incorporates a number of realistic factors that are often ignored in the literature. This leads to a model that accurately depicts the operations of college counseling centers. To demonstrate the benefits of the simulation model,  we use data from TAMU CAPS and perform a series of numerical experiments to investigate the impact of certain factors on the system's performance. Our experiments lead to key observations on the effect of different policy changes on the system-level resources at counseling centers. Firstly, increasing the proportion of external referrals or imposing maximum session limits do not result in the desired impact if the structure of the schedule topology is not considered. This is an important observation as many CAPS facilities have implemented such policies with the hope of improving the performance of the system. Secondly, our experiments on the schedule topology reveals that the proportions dedicated to each of the service types have a significant impact on the system performance. This is especially true when the system is overloaded with patients. Our results can be used by CAPS directors to identify the proportions for each of the service types that lead to the desired system performance. For the specific case of TAMU CAPS, it was identified that a first-time proportion of $12.5\%$ and a crisis proportion of $9.3\%$ would result in a system that meets their desired performance. It is important to note that such proportions will negatively impact the time to next session. This negative effect, however, can be mitigated by either increasing the proportion of external referrals or imposing a maximum session limit.

This work can be extended in several important directions. For example, a more rigorous investigation of the composition of optimal schedule topologies could result in even better system performance. This is valuable as our numerical experiments reveal, even for fixed proportions, the substantial impact of the topology on the performance of the system. Part of this analysis would include the investigation of time-varying schedule topologies in which the proportion dedicated to each service type can potentially change over time. This is of particular relevance to college counseling centers because of the predictable cyclical nature of the demand. Another interesting research direction is to enrich the simulation with group-based treatment options. These have been recently introduced by some CAPS facilities to handle the surge in demand. While group-based treatment options have the potential to improve system performance, their addition also results in a number of scheduling challenges. Within this context, the considered simulation model can be used to assess and quantify the impact of such options on the system. Lastly, this paper focuses on addressing the resource-level challenges facing CAPS. Another interesting dimension of the problem is the patient-level challenges such as the effectiveness of treatment plans. This is especially important when considering some of the policies being implemented by CAPS facilities. For example, does imposing a maximum session limit impact the effectiveness of treatments? Are group-based treatment options as effective as individual therapy? Establishing a quantitative framework to address these questions, in conjunction with the considered simulation model, can result in a counseling system with both a desirable system performance and patient outcome. We hope that this work can lead to further research in the aforementioned directions and potentially aid counseling centers in developing data-driven policies. 

\singlespacing
\printbibliography
%  \bibliographystyle{elsarticle-num} 
%  \bibliography{Refs}
% \bibliographystyle{plain}
% \bibliography{Refs}
\end{document}